\documentclass[pra,twocolumn,showpacs]{revtex4}
\usepackage{graphicx}
\usepackage{amsmath}

\newcommand{\re}[1]{\mathrm{Re}\,#1}
\newcommand{\im}[1]{\mathrm{Im}\,#1}
\renewcommand{\imath}[0]{\mathrm{i}}
\newcommand{\msum}{\sideset{}{'}\sum_{n=0}^{\infty}}

\begin{document}

 \title{Temperature dependence of the magnetic Casimir-Polder interaction}
 \author{H. Haakh, F. Intravaia, and C. Henkel}%
 \affiliation{
Institut f\"{u}r Physik und Astronomie, 
Universit\"{a}t Potsdam, Karl-Liebknecht-Stra{\ss}e 24/25, D-14476 Potsdam, Germany
}
 \author{S. Spagnolo and R. Passante}%
 \affiliation{
CNISM and Dipartimento di Scienze Fisiche ed Astronomiche,
Universit\`{a} degli Studi di Palermo, Via Archirafi 36, I-90123 Palermo, Italy
}

 \author{B. Power and F. Sols}%
 \affiliation{
Departamento de F\'isica de Materiales, Facultad de Ciencias F\'isicas,
Universidad Complutense de Madrid, Plaza de las Ciencias 1, E-28040 Madrid, Spain
}

\date{16 Oct 2009}


\begin{abstract}
We analyze the magnetic dipole contribution to atom-surface dispersion forces.
Unlike its electrical counterpart, it involves small transition frequencies
that are comparable to thermal energy scales. A significant temperature dependence
is found near surfaces with a nonzero dc conductivity, leading to a strong
suppression of the dispersion force at $T > 0$. We use thermal response theory 
for the surface material and discuss both normal metals and superconductors.
The asymptotes of the free energy of interaction and of the entropy are calculated 
analytically over a large range of distances. Near a superconductor, the onset 
of dissipation at the phase transition strongly changes the interaction, including 
a discontinuous entropy. We discuss the similarities with the Casimir interaction 
between two surfaces and suggest that precision measurements of the atom-surface interaction may shed light upon open questions around the temperature dependence of dispersion forces between lossy media.
\end{abstract}


\maketitle
\section{Introduction}
Ever since the work of Lennard-Jones \cite{Lennard-Jones} has the interaction between atoms and surfaces been of interest in many fields of physics, chemistry and technology.
The seminal work by Casimir and Polder \cite{Casimir48a} demonstrated that the shift in atomic energy levels close to a conductor is a probe for the quantum fluctuations of the electromagnetic field, a key concept of quantum electrodynamics (QED).
In this context, a nonzero temperature becomes relevant for several aspects
of the atom-surface interaction: thermally excited motion of
the surface (phonons) and inelastic scattering of atomic beams 
\cite{Zangwill,Desjonqueres, Farias}, occupation of excited atomic energy
levels, and 
enhancement of field fluctuations due to thermal photons 
\cite{Barton72}. The latter aspect is usually associated with distances
from the surface larger than the thermal wavelength $\Lambda_T =
\hbar c / 4 \pi k_B T$, approximately $0.6 \mu{\rm m}$ at room 
temperature. The free energy of interaction typically shows a change 
in power law with distance around this point: generally, it is enhanced with respect to zero temperature and becomes proportional to $T$.
This is the classical limit where the interaction is mainly
entropic in character \cite{Balian77,Balian78,Feinberg01}.

Experimental progress in recent years has achieved the sensitivity
required to detect the small energy shifts that occur at distances
on the order of $\Lambda_T$, making use of the exquisite control over the motion of atomic beams (beam deflection \cite{Sandoghar, Sukenik}, quantum reflection \cite{Shimizu, Druzhinina}) or clouds of ultracold laser-cooled atoms \cite{ Yuju2000, Obrecht07}.
The latter can be handled precisely in miniaturized traps implemented near solid state surfaces known as \emph{atom chips} \cite{Folman, Hansel01, Fortagh}.
These devices use optical or magnetic fields for trapping and can hold atomic clouds at distances down to a few microns. Here, the atom-surface interaction manifests itself typically as a distortion of the trapping potential (loss of atoms through tunneling to the surface or change in the trap oscillation frequency).
Therefore, the design of such setups requires exact knowledge of atom-surface interactions and conversely, theory predictions can be tested experimentally with high precision.  

A surprising result of the research on atom chips is the importance of magnetic field fluctuations near the surface arising from thermally excited currents in the material of the chip (Johnson noise related to Ohmic dissipation).
These fluctuations couple to the spin magnetic moment of the trapped atoms and are known to provoke the loss of atoms from the trap by flipping the sign of the potential \cite{Henkel99, Zhang05}. These losses are a main obstacle for technical applications and further down-scaling of atom chips.
Predictions that superconducting materials reduce the spin-flip induced losses significantly have recently been backed by trap lifetime measurements \cite{Skagerstam2006, Hohenester2007, Dikovsky2009, Emmert2009, Hufnagel2009, Kasch2009}.

In this paper, we address the magnetic dipole contribution to the atom-surface (Casimir-Polder) interaction including nonzero temperature. One would expect this to be a small correction to the electric dipole coupling \cite{McLachlan63a, Wylie85, Milonni94, Sols82, Buhmann07a, Bezerra08} because of the smallness of the transition matrix elements \cite{Bimonte2009a, Skagerstam09}. Yet, the strong magnetic mode density close to a metallic surface \cite{Jackson75, Joulain03, Henkel05a} and experimental evidence for magnetic spin flips call for a reconsideration of the magnetic contribution. In addition, the thermal occupation of photonic modes is quite relevant because magnetic transitions occur at much lower frequencies than electric ones, leading to a stronger temperature dependence.
Finally, it is well-known that dispersion forces between dielectric and magnetic materials can be repulsive  \cite{daSilva01}, as has been shown for the magnetic Casimir-Polder interaction at $T=0$ in Ref.\cite{Henkel05a}. We were thus led to investigate whether at distances beyond $\Lambda_T$ the total atom-surface interaction might be reduced due to the magnetic contribution. 

In this work, we calculate the magnetic Casimir-Polder free energy of interaction at different temperatures and consider a few well-known models for the electromagnetic response of the surface. Since Ohmic losses are crucial for the thermal behavior, it is highly interesting to compare both normal metals and superconductors. The latter are described here in the frame of the two-fluid model and Bardeen-Cooper-Schrieffer (BCS) theory \cite{Bardeen57,Schrieffer99}.
We demonstrate that the magnetic atom-surface coupling has very peculiar features unknown from its electrical equivalent. We find that for normal conductors at nonzero temperature, the magnetic dipole contribution to the  interaction is reduced, while it is enhanced for superconductors and in certain non-equilibrium situations. 
This resembles the macroscopic Casimir interaction between two dissipative plates, where the correct calculation of the force at large distances and nonzero temperatures has been the subject of debate \cite{Bostrom2000, Milton09a, Klimchitskaya09}.

This article is organized as follows.
In Sec. \ref{Atom-Surface interaction}, we give a brief review of the formalism used to calculate atom-surface interactions.
Section \ref{Polarizabilities and Green Tensors} presents the specific forms of the response functions and the experimental setups they describe. We also give expressions for the Green's tensor in different asymptotic regimes of the distance between the atom and the surface.
The magnetic Casimir-Polder free energy and entropy of an atom near metallic or superconducting surfaces at zero temperature is calculated in Sec. \ref{Magnetic Casimir-Polder}.
Section \ref{Casimir-Polder interaction at nonzero temperature} covers the effects at nonzero temperature and discusses the dissipative reduction in the interaction and questions connected to the entropy.
Non-thermal (out-of-equilibrium) states of the atoms that occur typically in experimental setups are investigated in Section \ref{Non-thermal states}.
We conclude summarizing and discussing the main results. 
Further technical details are given in the appendices.

\section{Atom-surface interaction}
\label{Atom-Surface interaction}
Quite a lot of research has been done on the 
interaction between an atom and a surface \cite{Casimir48a, McLachlan63a, Wylie85, Milonni94, Sols82, Buhmann07a, Bezerra08, Bimonte2009a, Skagerstam09}.
It can been shown from perturbation theory with respect to the
multipolar atom-field coupling \cite{McLachlan63a} that
the free energy of a polarizable
particle at nonzero temperature $T$ has the following general form
 (Einstein summation convention)
\begin{equation}
\mathcal{F}= -\frac{\hbar}{2\pi}
\int\limits_{0}^{\infty}{\rm d}\omega
\coth\!\left( \frac{\hbar\omega}{2k_{B}T} \right)
\im\![\beta^{T}_{ij}(\omega) \mathcal{H}_{ji}(L, \omega)]
.
\label{eq1}
\end{equation}
Here, $\boldsymbol{\beta}^{T}$ is the (magnetic or electric) polarizability 
tensor for the atom in a thermal state of temperature $T$, 
and $\boldsymbol{\mathcal{H}}$ is the (corresponding) Green's tensor in the presence of the surface, defined in Eq.(\ref{eq:def-induced-field}) below.
In the planar geometry we are interested in, the Green's tensor depends only on the atom-surface distance $L$ and on frequency. It is well known that Eq.(\ref{eq1}) has the same form for electric or magnetic dipole couplings \cite{Buhmann07a,Boyer74}; our notation 
is adapted to the magnetic case. 
A simple and general derivation of Eq.\eqref{eq1} can be given following Refs.\cite{Henkel2002,Novotny08}.
The effective interaction potential between a polarizable particle and the 
(magnetic or electric) field $\boldsymbol{B}$ \cite{Jackson75} is given by
\begin{equation}
\mathcal{F}=
- \frac{\langle \boldsymbol{\mu} \cdot \boldsymbol{B}(\boldsymbol{r}_{0})\rangle_{T}}{2}.
\label{eneq}
\end{equation}
The expectation value $\langle \cdots\rangle_T$ is taken in an equilibrium state 
of the non-coupled system at temperature $T$ and implicitly evaluates 
symmetrically ordered operator products;
$\boldsymbol{\mu}$ is the (magnetic or electric) dipole operator 
and $\boldsymbol{B}$ the corresponding
field operator, evaluated at the atom position $\boldsymbol{r}_{0}$.
The factor $1/2$ arises from a coupling constant integration (excluding a permanently polarized atom). 
Within first-order perturbation theory, both the dipole moment $\boldsymbol{\mu}(t)$ and 
the field $\boldsymbol{B}(t)$
can be split into fluctuating (fl) and induced (in) parts: the fluctuating part describes the 
intrinsic equilibrium fluctuation, while the induced part arises in perturbation theory from
the dipole coupling \cite{Mandel95}.
Eq.\eqref{eneq} becomes
\begin{equation}
\mathcal{F}=-\frac{\langle \boldsymbol{\mu}^{\mathrm{in}}(t) 
\cdot \boldsymbol{B}^{\mathrm{fl}}(\boldsymbol{r}_{0},t)\rangle_{T}}{2}
- \frac{\langle \boldsymbol{\mu}^{\mathrm{fl}}(t) 
\cdot \boldsymbol{B}^{\mathrm{in}}(\boldsymbol{r}_{0},t)\rangle_{T}}{2}
	\label{eq:F-split-in-fl}
\end{equation}
Here, we assume the fluctuating parts of the dipole and of the field to be 
decorrelated at this order. This assumption would break down at higher orders
of perturbation theory. Note that while in Eq.(\ref{eneq}), the total dipole and 
field operators (Heisenberg picture) commute at equal times, this is no 
longer true for their
`in' and `fl' constituents in Eq.(\ref{eq:F-split-in-fl}).
The induced quantities are given, in frequency space,
by the retarded response functions \cite{Jackson75}
\begin{eqnarray}
&\mu_{i}^{\mathrm{in}}(\omega) = \beta_{i j}(\omega) B_j^{\mathrm{fl}}(\boldsymbol{r}_{0},\omega) 
\label{eq:def-induced-field}
\\
&B_i^{\mathrm{in}}(\boldsymbol{r},\omega) = 
\mathcal{H}_{ij}(\boldsymbol{r},\boldsymbol{r}_{0},\omega)
\mu_j^{\mathrm{fl}}(\omega)~,
\nonumber\label{eq6}
\end{eqnarray}
where the frequency dependence allows for a temporal delay. 
The equilibrium fluctuations follow from the fluctuation-dissipation 
theorem \cite{Callen51}
\begin{eqnarray}
\langle B_i^{\mathrm{fl}}(\boldsymbol{r},\omega)
	B_j^{\mathrm{fl}}(\boldsymbol{r},\omega') \rangle_T
&=& \frac{\hbar}{2\pi } \delta(\omega-\omega')
\coth\left(\frac{\hbar \omega}{2k_B T}\right)
\nonumber\\
&& \times 
\im\![\mathcal{H}_{i j}(\boldsymbol{r}, \boldsymbol{r}, \omega)]
~,
	\label{eq:FTD_field}
\\
\langle \mu_i^{\mathrm{fl}}(\omega)\mu_j^{\mathrm{fl}}(\omega')\rangle_T
&=& \frac{\hbar}{2\pi }\delta(\omega-\omega')
\coth\left(\frac{\hbar \omega}{2 k_B T}\right)
\nonumber\\
&& \times
\im\![ \beta_{i j}^{T}(\omega)]~.
	\label{eq:FTD_dipole}
\end{eqnarray}
Combining Eqs.(\ref{eq:F-split-in-fl}--\ref{eq:FTD_dipole}), 
we recover Eq.\eqref{eq1}, setting 
$\mathcal{H}_{i j}(\boldsymbol{r}_0, \boldsymbol{r}_0, \omega) = 
\mathcal{H}_{i j}(L, \omega)$.
One uses the fact that the imaginary part of both Green's tensor and 
polarizability tensor are odd in $\omega$ (retarded response functions). 
The field correlations are needed at the same position $\boldsymbol{r}_0$,
and it is easy to remove the divergent free-space contribution (Lamb shift)
from $\mathcal{F}$, by keeping in the Green's tensor only the reflected part 
of the field \cite{Wylie85}. 
In a planar geometry, it follows from symmetry
that the result can only depend on the dipole-surface distance $L$. Note
that the Green's tensor can also depend on temperature through the surface 
reflectivity. In a two-level model for the atom, the thermal polarizability 
$\beta_{i j}^{T}(\omega)$ contains a stronger $T$-dependence because of
a Fermi-Dirac-like statistics \cite{Callen51}, see Eq.(\ref{eq:two-level_polarizability}) 
below.

%
Eq.\eqref{eq1} is often expressed in an equivalent form using the analyticity of
$\boldsymbol{\beta}^{T}(\omega)$ 
and $\boldsymbol{\mathcal{H}}(L, \omega)$ in the upper half of
the complex frequency plane.
Performing a rotation onto the imaginary frequency axis yields the so-called Matsubara expansion \cite{Matsubara55}
\begin{equation}
\label{eq:Matsubara-series}
\mathcal{F}(L, T)=-k_{B}T
\msum
\beta^{T}_{i j}(\imath \xi_n)
\mathcal{H}_{j i}
(L, \imath \xi_n)~,
\end{equation}
where $\xi_{n}=2\pi n k_{B}T/\hbar$ are the Matsubara frequencies and the prime in the sum indicates that the $n=0$ term must be weighted by a prefactor $1/2$.
Both $\boldsymbol{\beta}^{T}(\imath \xi)$ and $\boldsymbol{\mathcal{H}}(L, \imath \xi)$ are real expressions for $\xi>0$. 

If the atom is in a well defined state $|a\rangle$ rather than in a 
thermal mixture,
we have the expression of Wylie and Sipe \cite{Wylie85}
\begin{multline}
\label{WS}
\mathcal{F}(L, T)=-k_{B}T \msum \beta^{a}_{ij}(\imath \xi_n)\mathcal{H}_{ji}(L,\imath \xi_n)\\
+\sum_b n(\omega_{ba}) \mu_i^{ab} \mu_j^{ba}
\re [\mathcal{H}_{ji}(L,\omega_{ba})]~,
\end{multline}
where $\boldsymbol{\beta}^a$ is the state-specific polarizability 
\cite{McLachlan63,Wylie85} 
\begin{equation}
\beta_{ij}^a({\omega})=\sum_b
\frac{\mu_i^{ab}\mu_j^{ba}}{\hbar}\frac{2\omega_{ba}}{\omega_{ba}^{2}-(\omega+\imath 0^{+})^{2}}
\label{eq:WS_polarizability}
\end{equation}
Here,
$\mu_i^{ab}=\langle a| \mu_{i} | b\rangle$ are the dipole matrix elements, 
$\omega_{ba}$ is the frequency of the 
virtual transition $|a \rangle \rightarrow |b\rangle$ ($\omega_{ba} < 0$ for
a transition to a state of lower energy). Finally, the thermal occupation of photon modes %
\begin{equation}
n(\omega)=\left(e^{\hbar\omega / k_{B}T }-1\right)^{-1}~
\end{equation}
%
 in the second term in Eq.(\ref{WS}) is the Bose-Einstein distribution.
At $T=0$ it occurs only for excited states, for which $n(\omega_{ba}) \to -1$ for $\omega_{ba} < 0$
(see Eqs.(4.3, 4.4) of Ref.\cite{Wylie85}). The real
part of the Green's tensor can be given an interpretation 
from the radiation reaction of a classical dipole oscillator \cite{Wylie85}.
Similarly, this term
is practically absent for the electric Casimir-Polder
interaction of ground-state atoms because of the higher transition frequencies,
$\Omega_e \approx (k_B / \hbar)\,10^{3} \ldots 10^4 \,\mathrm{K}$.

In the following, we call the Matsubara sum (first line) in Eq.(\ref{WS})
the \emph{non-resonant} contribution, and the second line the \emph{resonant} 
one, because it involves the field response at the atomic transition frequency.

\section{Response functions}
\label{Polarizabilities and Green Tensors}

The formalism presented in the previous section is quite general and 
$\boldsymbol \beta$ [$\boldsymbol{\mathcal{H}}$] could represent either 
the magnetic or electric polarizability [Green's tensor], respectively. We now give 
the specific forms of these quantities in the magnetic case, focusing on a 
planar surface and specific trapping scenarios. 

\subsection{Green's tensors and material response}
\label{Materials}
The Green's tensor for a planar surface can be calculated analytically.
Let the atom be on the positive $z$-axis at a distance $L$ from a medium
occupying the half-space below the $xy$-plane.
By symmetry, the magnetic Green's tensor 
$\boldsymbol{\mathcal{H}}(L, \omega ) = 
\boldsymbol{\mathcal{H}}(\boldsymbol{r}_0, \boldsymbol{r}_0, \omega )$ 
is diagonal and invariant under rotations in the $xy$-plane:
\begin{widetext}
\begin{equation}
\boldsymbol{\mathcal{H}}(L,\omega) = \frac{\mu_0}{8 \pi}
\int\limits_0^\infty k dk \, \kappa
\left[\left(r^{\rm TE}(\omega, k)+
 \frac{\omega^2}{c^2 \kappa^{2}}r^{\rm TM}(\omega, k)\right)[\boldsymbol{\hat x\hat x}+ \boldsymbol{\hat y\hat y}]+ 2\frac{k^{2}}{\kappa^{2}}r^{\rm TE}(\omega, k) \boldsymbol{\hat z\hat z}\right]e^{-2 \kappa L }~,
\label{eq:magnetic_GF}
\end{equation}
\end{widetext}
where $\mu_{0}$ is the vacuum permeability 
and $\boldsymbol {\hat x\hat x}$, $\boldsymbol{\hat y\hat y}$ and $\boldsymbol{\hat z\hat z}$ are the Cartesian dyadic products. 
We consider here a local isotropic, nonmagnetic bulk medium 
[$\mu(\omega)=1$], so that the Fresnel formulae give
the following reflection coefficients in the TE- and TM-polarization 
(also known as $s$- and $p$-polarization) \cite{Jackson75}
\begin{equation}
r^{\rm TE}(\omega, k)=\frac{\kappa-\kappa_{m}}{\kappa+\kappa_{m}},
\quad
r^{\rm TM}(\omega, k)=\frac{\epsilon(\omega)\kappa-\kappa_{m}}{\epsilon(\omega)\kappa+\kappa_{m}}~,
\label{eq:fresnel}
\end{equation}
where $\kappa$, $\kappa_m$ are the propagation constants in vacuum
and in the medium, respectively (roots with $\im{\kappa}\le 0; \re{\kappa} \ge 0$)
\begin{equation}
\label{eq:kappa}
\kappa=\sqrt{k^2-\frac{\omega^2}{c^2}},\quad 
\kappa_{m}=\sqrt{k^2-\epsilon(\omega)\frac{\omega^2}{c^2}},
\end{equation}
and $k=|\boldsymbol{k}|$ is the modulus of the in-plane wavevector.
Note that the magnetic Green's tensor can be obtained from the electric one 
$\boldsymbol{\mathcal{G}}$ by 
swapping the reflection coefficients \cite{Henkel99}
\begin{equation}
\label{eq:electric_GF}
\boldsymbol{\mathcal{H}}\equiv c^{-2}\boldsymbol{\mathcal{G}}(r^{\rm TE}\leftrightarrow r^{\rm TM})
\end{equation}

All information about the optical properties of the surface is encoded 
in the dielectric function $\varepsilon(\omega)$.
We will use four different commonly established descriptions, each of which includes Ohmic dissipation in a very characteristic way. As it turns out, the magnetic
Casimir-Polder interaction is much more sensitive to dissipation than the
electric one (see Sec.\ref{Dissipative quenching}). This is due to the fact 
that the resonance frequencies in
the magnetic polarizability $\boldsymbol{\beta}(\omega)$ are much lower
(see Sec.\ref{s:beta-and-alpha}).

The first model is a \emph{Drude metal} \cite{Jackson75}
\begin{equation}
\varepsilon_\mathrm{Dr}(\omega) =1-\frac{\omega_{p}^{2}}{\omega(\omega+\imath\gamma)}~,
	\label{eq:def-Drude-epsilon}
\end{equation}
where $\omega_{p}$ is the plasma frequency and $\gamma > 0$ 
is a phenomenological dissipation rate.
This is the simplest model for a metal with finite conductivity. 
If $\gamma$ is constant (independent of temperature), the
conductivity can be attributed to impurity scattering in the medium.

The second model is the dissipationless \emph{plasma model} 
$\varepsilon_\mathrm{pl}(\omega)$: here, one sets $\gamma = 0$
in the right-hand side of Eq.(\ref{eq:def-Drude-epsilon}). This corresponds to a purely
imaginary conductivity.

In the context of atom chips, the case of a \emph{superconductor} is
particularly interesting because dissipation is suppressed as the temperature
$T$ drops below the critical temperature $T_c$.  We adopt here (third
model) a 
description in terms of the two-fluid model, a weighted sum of a 
dissipationless supercurrent response (plasma model) and a normal conductor
response 
\begin{eqnarray}
\varepsilon_\mathrm{sc}(\omega, T) &=& \eta(T) \varepsilon_\mathrm{pl}(\omega) + [1-\eta(T)] \varepsilon_\mathrm{Dr}(\omega)~,\\
\eta(T) &=& \left[ 1 - \left(\frac{T}{T_c}\right)^4\right] \Theta(T_c -T)
~,
	\label{eq:epsilon-two-fluid-model}
\end{eqnarray}
where the \emph{order parameter} $\eta(T)$ follows the Gorter--Casimir 
rule \cite{Schrieffer99}.
At $T=0$, the superconductor coincides with the plasma model, as is
known from the London theory of superconductivity \cite{London35}.
The plasma model is thus the simplest description of a superconductor at zero 
temperature rather than a model for a normal metal.
More involved descriptions of superconductors (including BCS theory)
also reproduce the plasma behavior at low frequencies ($\omega$ well below 
the BCS gap) and temperature close to absolute zero.
The full BCS theory of superconductivity can be applied in this context, too,
using its optical conductivity \cite{Mattis58, Zimmermann91, Berlinsky93},
as recently discussed in Ref.\cite{BCSunpub}. We shall see below
(Sec.\ref{sec:superconductor}), however, that the two-fluid model and
BCS theory give very close results for realistic choices of the physical 
parameters.

Our fourth model takes a look at the peculiar case of a very clean metal.
Here, rather than by impurity scattering, dissipation is dominated by electron-electron or electron-phonon scattering. In these cases, the dissipation rate 
in the Drude formula~(\ref{eq:def-Drude-epsilon})
follows a characteristic power law 
\begin{equation}\label{eq:perfect_crystal}
\gamma(T) \propto T^n, \qquad n>1
\end{equation}
at small temperatures and saturates to a constant value at high temperatures 
(Bloch-Gr\"uneisen law).
It is reasonable to call this system the \emph{perfect crystal} model.
As in a superconductor, dissipation is turned on by temperature, but in a completely different manner. This can be distinguished in the atom-surface
interaction potential.
\subsection{Distance dependence of the Green's tensors}
\label{s:distance-regimes}
For the Drude model,
there are three different regimes for the atom-surface distance that are
determined by physical length scales of the system
(see Ref.\cite{Henkel05a} for a review):
the skin depth in the medium,
\begin{equation}
	\delta_\omega = \frac{\lambda_{\rm p}}{2\pi}\sqrt{\frac{2\gamma}{\omega}}
	,
	\label{eq:def-skin-depth}
\end{equation}
where $\lambda_{\rm p} = 2 \pi c / \omega_p$ is the plasma wavelength,
and the photon wavelength in vacuum, 
\begin{equation}
	\lambda_\omega = \frac{ 2\pi c }{ \omega }
	.
	\label{eq:def-photon-wavelength}
\end{equation}
Note that $\varepsilon( \omega ) \approx 2{\rm i}\lambda_\omega^2/(2\pi\delta_\omega)^2$
for frequencies $\omega \ll \gamma \ll \omega_p$ (Hagen-Rubens regime).
This is the relevant regime for the relatively low magnetic resonance 
frequencies. 
We then have $\delta_\omega \ll \lambda_\omega$ which
leads to the following three domains:
(i) the \textit{sub-skin-depth region}, $L\ll \delta_\omega$,
(ii) the 
\textit{non-retarded region}, $ \delta_\omega \ll L\ll \lambda_\omega$, and
(iii) the \textit{retarded region}: $\lambda_\omega \ll L$.
In zones (i) and (ii), retardation can be neglected (van-der-Waals zone),
while in zone (iii), it leads to a different power law (Casimir-Polder zone) for
the atom-surface interaction.

Since the boundaries of the three distance zones depend on frequency, the respective length scales differ by orders magnitude between the magnetic and the electric case.
For electric dipole transitions, the Hagen--Rubens regime cannot be applied
because the resonant photon wavelength is much smaller.
The role of 
the skin depth is then taken by the plasma wavelength $\lambda_{\rm p}$.
This implies
that the Casimir-Polder zone (iii) for the electric dipole interaction
occurs in a range of distances where magnetic retardation is still negligible
[zones (i) and (ii)].


In the three regimes, different approximations for the reflection coefficients 
that appear in the Green's function~(\ref{eq:magnetic_GF})
can be made. We start with the Drude model where
in the \textit{sub-skin-depth zone} \cite{Henkel99}, we have $k\gg 
1/\delta \gg 1/\lambda$ and
%
\begin{eqnarray}
r^{\rm TE}(\omega, k) &\approx& [\epsilon(\omega)-1]\frac{\omega^2}{4c^2k^2},\nonumber\\
r^{\rm TM}(\omega, k)&\approx& \frac{\epsilon(\omega)-1}{\epsilon(\omega)+1}\left[1+
\frac{\epsilon(\omega)}{\epsilon(\omega)+1}\frac{\omega^2}{c^2k^2}\right].
\label{eq:r_subskindepth}
\end{eqnarray}
At intermediate distances in the \textit{non-retarded zone},
the wavevector is $1 / \lambda \ll k \ll 1 / \delta$, hence,
\begin{eqnarray}
r^{\rm TE}(\omega, k) &\approx& -1+\imath\frac{2}{\sqrt{\epsilon(\omega)}}\frac{c k}{\omega},\nonumber\\
r^{\rm TM}(\omega, k) &\approx& 
1+\imath\frac{2}{\sqrt{\epsilon(\omega)}}\frac{\omega}{c k}.
	\label{eq:r-non-retarded}
\end{eqnarray}
Finally, in the  \textit{retarded zone} we can consider $k\ll 1 / \lambda \ll 1/ \delta$, so that 
\begin{eqnarray}
r^{\rm TE}(\omega, k) &\approx& -1+\frac{2}{\sqrt{\epsilon(\omega)}},\nonumber\\
r^{\rm TM}(\omega, k) &\approx& 1-\frac{2}{\sqrt{\epsilon(\omega)}}.
\label{eq:epsilon_retarded}
\end{eqnarray}
A similar asymptotic analysis can be performed for the other model dielectric
functions. It turns out that Eqs.(\ref{eq:r_subskindepth}--\ref{eq:epsilon_retarded})
can still be used, provided the assumption $|\epsilon(\omega)| \gg 1$ holds.
This is indeed the case for a typical atomic magnetic dipole moment and
a conducting surface. 

The asymptotics of the Green's functions that correspond to these
distance regimes 
are obtained by performing the $k$-integration in Eq.(\ref{eq:magnetic_GF})
with the above approximations for the reflection coefficients. The results
are collected in Table \ref{MagneticAppr}. One 
notes that the $zz$-component is larger by a factor $2$ compared to
the $xx$- and $yy$-components. The difference
between the normal and parallel dipoles can be understood by the
method of images \cite{Jackson75}.
Furthermore, the magnetic response for
a normally conducting metal in the sub-skin-depth regime is purely
imaginary and scales linearly with the frequency $\omega$
-- the reflected magnetic field is generated by induction. 
Only the superconductor or the plasma model can reproduce
a significant low-frequency magnetic response, via the Mei\ss{}ner-Ochsenfeld effect.
In contrast, the electric response is strong for all conductors because surface charges screen the electric field efficiently.

The imaginary part of the Green's functions determines the local mode
density (per frequency) for the magnetic or electric 
fields \cite{Joulain03}. These can be compared directly after 
multiplying by $1/\mu_0$ (or $\varepsilon_0$), respectively.
As is discussed in Refs.\cite{Joulain03, Henkel05a}, in the sub-skin-depth
regime near a metallic surface, the field fluctuations are mainly of 
magnetic nature. This can be traced back to surface charge screening 
that efficiently decouples the interior of the metal and the vacuum above.
Magnetic fields, however, cross the surface much more easily as surface 
currents are absent (except for superconductors). This reveals,
in the vacuum outside the metal, the thermally 
excited currents from the bulk.

\begin{table*}[tbh]
\centering
\begin{tabular}{l| c c c c c |c}
\hline\hline
& \hspace{.3cm}Sub -&&\hspace{-.8cm}skin depth&\vline&Non-retarded & Retarded\\[1ex]
&Drude &\vline &plasma&\vline&\\
\hline
$\mathcal{H}_{xx}$ 
&$\displaystyle \frac{\imath \mu_0}{32 \pi \delta^2_\omega L}$
&\vline
& $\displaystyle - \frac{\mu_0  \pi}{16 \lambda_p^2 L}$ 
&\vline
& $\displaystyle - \frac{ \mu_0}{32 \pi L^3}$
& $\displaystyle - \frac{ \mu_0}{32 \pi L^3} \left(1 - \frac{2 i \omega L}{c} - \frac{4 \omega^2 L^2}{c^2}\right)e^{2 i \omega L / c}$\\[1ex]
$\mathcal{G}_{xx}$ 

& 
&
& $\displaystyle \frac{1}{32\pi\epsilon_0 L^3}$ 
&
&
& $\displaystyle \frac{1}{32\pi\epsilon_0 L^3}\left(1-\frac{2 \imath\omega L}{c}-\frac{4\omega^2 L^2}{c^2}\right)e^{2 \imath \omega L / c}$ \\
\hline
\end{tabular}
\\
\caption{Magnetic and electric Green's tensors at a planar surface.
The other elements have the asymptotes
$\mathcal{H}_{yy} = \mathcal{H}_{xx}$,
$\mathcal{H}_{zz} = 2\mathcal{H}_{xx}$, and similarly for
$\mathcal{G}_{ii}$. The off-diagonal elements vanish.}
\label{MagneticAppr}
\end{table*}

\subsection{Atomic polarizability}
\label{s:beta-and-alpha}

The magnetic and electric
polarizabilities are determined by the transition dipole matrix elements
and the resonance frequencies. 
We are interested in the retarded response function, which for an arbitrary atomic state $|a\rangle$ is given by Eq.(\ref{eq:WS_polarizability}) above.

When the atom is in thermal equilibrium, we have to sum the polarizability
over the states $|a\rangle$ with a Boltzmann weight:
\begin{equation}
\beta^{T}_{ij}(\omega) = \sum_{a}
\frac{e^{-E_{a} / k_B T}}{Z}
\beta^{a}_{ij}(\omega)
\label{eq:WS_polarizability_thermal}
\end{equation}
where $Z$ is the partition function. In the limit $T\rightarrow 0$, we recover the polarizability for a ground state atom.
%
For a two-level system with transition frequency $\Omega_m$, 
the previous expression takes a simple form and can be expressed 
in terms of the ground state polarizability [Eq. (\ref{eq:WS_polarizability}), where $a=g$]:
\begin{equation}\label{eq:two-level_polarizability}
\boldsymbol{\beta}^{T}(\omega) 
= 
\tanh\left(\frac{\hbar\Omega_m}{2k_B T}\right) \boldsymbol{\beta}^{g}(\omega)
~.
\end{equation}

Let us now compare the electric and magnetic polarizabilities.
The magnetic transition moment among states with zero orbital spin
scales with
$\mu_B g_s$ where $g_{s}$ is the Land\'{e} factor for the electron spin
and $\mu_B$ the Bohr magneton. Electric dipoles are
on the order $e a_0$ with $a_0$ the Bohr radius.
With the estimate that the resonance frequencies ($\Omega_m$ and 
$\Omega_e$) determine the relevant range
of frequencies, we have approximately
\begin{equation}
\label{eq:magnitude}
\frac{\alpha(0) / \varepsilon_0  }{ \beta(0)\mu_0 }
\frac{ \Omega_e }{ \Omega_ m }
\sim \frac{1}{\alpha^{2}_{\rm fs}} 
\end{equation}
where $\alpha_{\rm fs}\approx 1/137$ is the fine-structure constant. 
The magnetic interaction is thus expected to be a small correction.
Conversely, the narrower range of frequencies makes it much more
sensitive to the influence of temperature. 

We have seen now that the polarizability of an atom takes a positive constant value at low frequency and the induced magnetic dipole is parallel to the magnetic field (paramagnetism).
Let us consider for comparison a metallic nanosphere. If its radius $R$ is smaller than the penetration depth and the wavelength, the polarizability is given by \cite{Jackson75}
\begin{equation}
\label{betananosphere}
\beta_{\rm sph}(\omega)=\frac{2\pi}{15 \mu_0}\left(\frac{R\omega}{c}\right)^{2}[\varepsilon(\omega)-1]R^{3}~.
\end{equation}
This quantity vanishes at low frequencies and has a negative real part (diamagnetism).
For a qualitative comparison to an atom one can estimate, e.g., the magnetic oscillator strength, defined by the integral over the  imaginary part of the polarizability at real frequencies
\begin{eqnarray}
\int_0^\infty \beta(\omega) d\omega &=& \frac{\pi \mu_B^2}{\hbar}\\
\int_0^{\omega_p \approx \infty} \beta_{\rm sph}(\omega) d\omega &=& \frac{2 \pi}{15 \mu_0}  \gamma  \left(\frac{\omega_{ p}}{c}\right)^2 R^5 \log(\frac{\gamma}{\omega_p})~.
\end{eqnarray}
From the Clausius-Mossotti relation follows the electric counterpart:
\begin{eqnarray}
\int_0^\infty \alpha_{\rm sph}(\omega) d\omega &=& \frac{2\pi^2}{\sqrt{3} }\epsilon_0 \omega_{ p} R^3 + \mathcal{O}(\frac{\gamma}{\omega_p}).
\end{eqnarray}

We find that the nanoparticle has a dominantly electric response, similar to an atom, but the ratio of the oscillator strengths depends on the material parameters and the radius. 
From the above expressions, we find that the absolute value of the nanosphere's magnetic oscillator strength is actually smaller than the one of an atom if the sphere's radius $R \lesssim 1 {\rm nm}$.

\subsection{Optical and magnetic traps}
	\label{s:optical-magnetic-traps}

The resonance frequencies relevant for the magnetic Casimir-Polder
potential depend on the trapping scheme.
We focus here on alkali atoms that are typically used in ultracold
gases and distinguish between optical and magnetic traps.

In an \emph{optical trap}, we may consider the case that the magnetic 
sublevels are degenerate
and subject to the same trapping potential (proportional to the intensity
of a far-detuned laser beam). Magnetic dipole transitions can then occur 
between hyperfine levels whose splitting is on the order of 
$\Omega_{m} / 2\pi \approx 10^8, \ldots, 10^{10}\, \mathrm{Hz}$,
corresponding to temperatures of $5, \ldots, 500\,\mathrm{mK}$
(see Appendix~\ref{Magnetic transition matrix elements} for more details.)
In contrast, electric dipole transitions occur in the visible range 
$\Omega_e / 2\pi \approx 10^{15}\mathrm{Hz}$ or 
$\sim 50\,000\,\mathrm{K}$. If we average over the magnetic sublevels,
we get an isotropic magnetic polarizability.
This allows to 
write $\beta^T_{ij}= \beta^T_{\rm iso} \frac{1}{3} \delta_{ij}$, 
so that in Eq.(\ref{eq1}) or~(\ref{eq:Matsubara-series})
\begin{equation}\label{eq:beta_iso}
\beta^T_{ij} \mathcal{H}_{ji}
= 
\beta^{T}_{\rm iso}
\frac{2 \mathcal{H}_{xx}  + \mathcal{H}_{zz}
} { 3}
\end{equation}
%

The setup we will consider in most of our examples is an atom in a 
\emph{magnetic trap}. In these traps, one uses the interaction of a permanent magnetic dipole with an inhomogeneous, static magnetic field $\boldsymbol{B}$.
Let us consider for simplicity a spin $1/2$ manifold:
the Zeeman effect then leads to a splitting of the magnetic sublevels 
by the Larmor frequency, $\Omega_{m} = \mu_B g_s |\boldsymbol{B}|/\hbar$
in weak fields.
To give an order of magnitude, $\Omega_m / 2 \pi \approx
280\,\mathrm{MHz} \approx (k_B / 2\pi\hbar) \,13.5\,\mathrm{mK}$ 
at $B = 10 \mathrm{mT}$.
Atoms in those magnetic sublevels where 
$\Delta E = - \boldsymbol{\mu} \cdot \boldsymbol{B} > 0$ are weak-field seekers, and can be trapped in field minima.
The magnetic trap we have in mind is a two-wire trap suspended below
the surface of an atom chip. Currents in the two wires, combined with a
static field, create a field minimum below the chip surface, with gravity pulling
the potential minimum into a position where the magnetic field is nonzero 
and perpendicular to the surface. Magnetic dipole transitions are then generated
by the parallel components $\mu_{x}$, $\mu_{y}$ of the dipole moment (see Appendix~\ref{Magnetic transition matrix elements}).
In this anisotropic scenario, the components of the magnetic polarizability tensor are given by 
$\beta^T_{xx}(\imath\xi) = \beta^T_{yy}(\imath\xi) 
= \beta^T_{\rm an}(\imath \xi)$ and $\beta^T_{zz}(\imath\xi)  = 0$.
The relevant components of the Green's tensor are, therefore,
\begin{equation}\label{eq:FE-aniso}
\beta^T_{ij} \mathcal{H}_{ji}
= 
\beta^{T}_{\rm an}\,
2 \mathcal{H}_{xx}
~.
\end{equation}
We should mention that many experiments do not realize a global equilibrium situation, as assumed in Eq.(\ref{eq1}).
In typical atom chip setups, atoms are laser cooled to $\mu\mathrm{K}$ 
temperatures or prepared in a well-defined state, while the surface is generally 
at a much higher temperature, even when superconducting.
For the description of such situations, a more general approach \textit{\`{a} la} Wylie and Sipe [Eqs.(\ref{WS}) and (\ref{eq:WS_polarizability})] is more 
suitable, and we discuss the results in Sec.\ref{Non-thermal states}. Before 
addressing these, we start with thermal equilibrium free energies, however.
This
may be not an unrealistic assumption in spectroscopic experiments where 
Casimir-Polder energies are measured with atoms near the window of a 
vapor cell \cite{Failache99}. From the theoretical viewpoint, thermal equilibrium
provides an unambiguous definition of the entropy related to the atom-surface
interaction. We shall see that this quantity shows remarkable features 
depending on the way dissipation and conductivity is included in the material
response. This closely parallels the issue of the thermal correction to the
macroscopic Casimir interaction, a subject of much interest lately.

To summarize, in an optical (isotropic) trap, the equilibrium Casimir-Polder
free energy~\eqref{eq:Matsubara-series} is given by the Matsubara sum
\begin{eqnarray}\label{eq:FE_iso}
&&
\mathcal{F}_\mathrm{iso}(L, T) = 
\\
&&
- k_B T \msum \beta^{T}_{\rm iso}(\imath \xi_n)
\frac{2 \mathcal{H}_{xx}(L, \imath \xi_n)  + \mathcal{H}_{zz}(L, \imath \xi_n)
} { 3}
\nonumber
~,
\end{eqnarray}
while in a magnetic (anisotropic) trap, we have
\begin{eqnarray}\label{eq:FE_aniso}
\mathcal{F}_\mathrm{an}(L, T) = 
- 2 k_B T \msum \beta^{T}_{\rm an}(\imath \xi_n)
{\mathcal{H}}_{xx}
(L,\imath \xi_n)~.
\end{eqnarray}

\section{Zero-temperature interaction potential}
\label{Magnetic Casimir-Polder}

\begin{figure}[hhh]
\centering
\includegraphics[width=8cm]{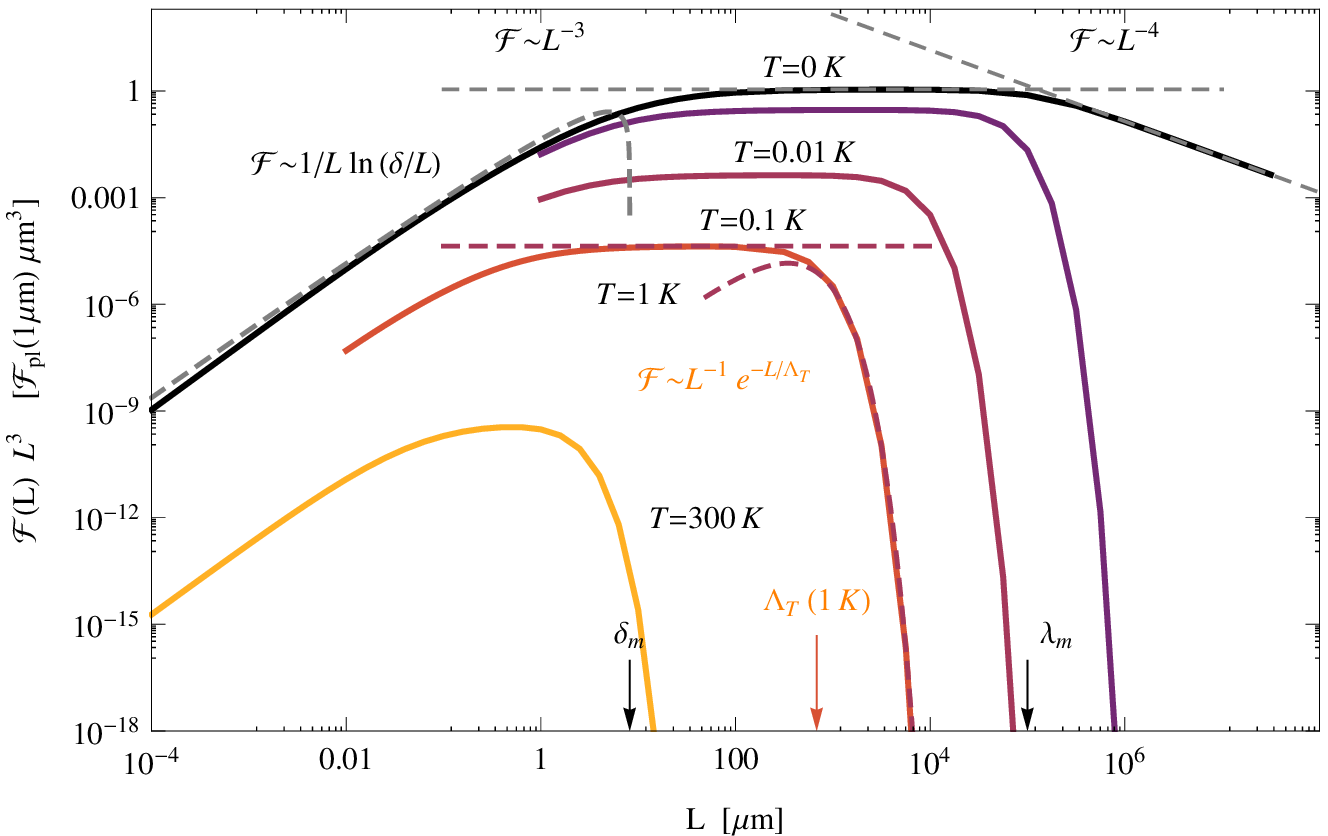}
\includegraphics[width=8cm]{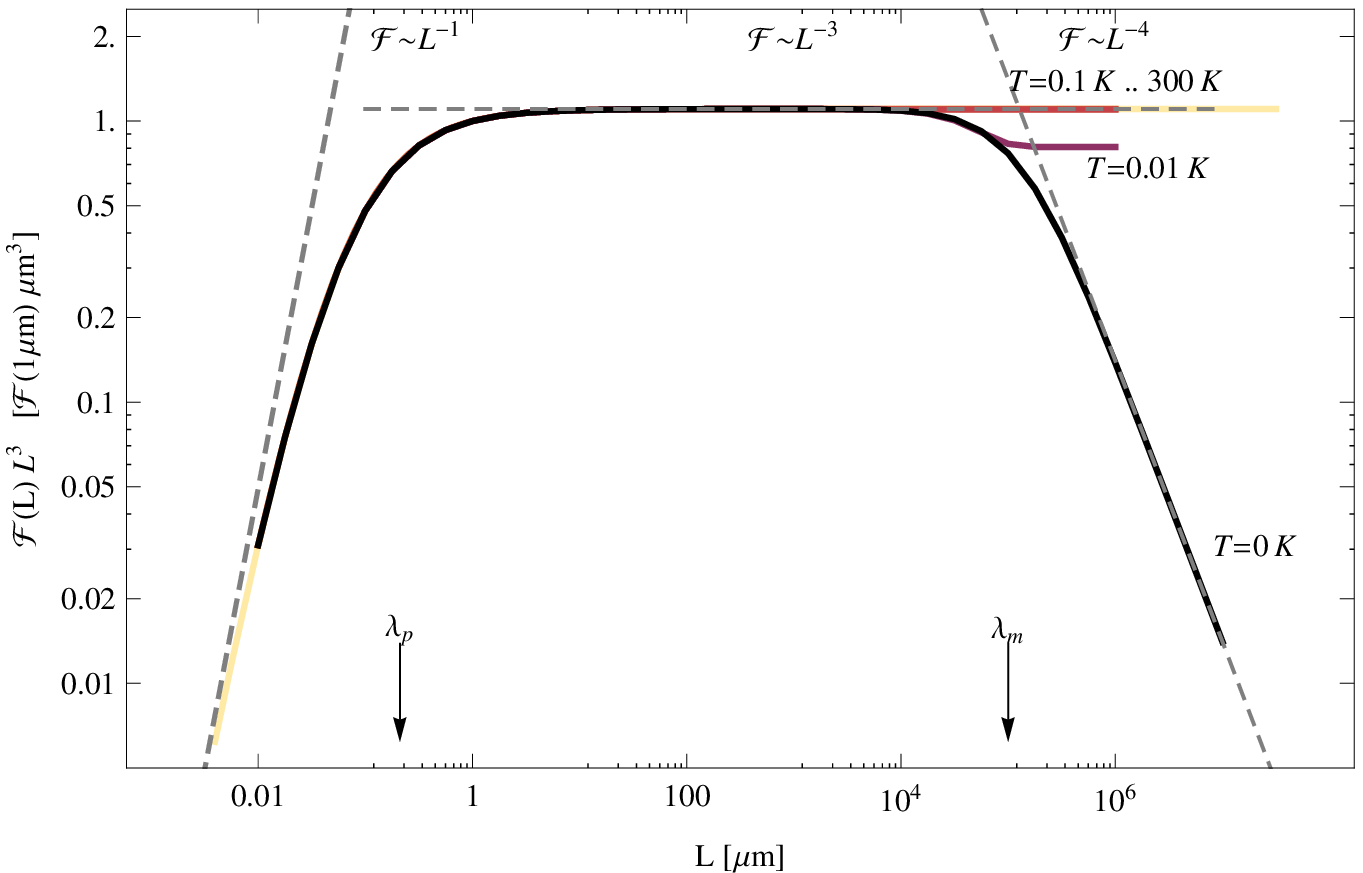}
\caption[$\mathcal{F}(L)$ anisotropic dipole, plasma and Drude model.]{(Color online) Casimir-Polder free energy vs.\ distance $L$ for an anisotropic magnetic dipole
(transition moment parallel to the surface). 
Drude model (Top) and plasma model (Bottom), with plasma frequency
$\omega_p / 2\pi \approx 1.42 \times 10^{15} \mathrm{Hz}$ 
($\lambda_{\rm p} \approx 210\,\mathrm{nm}$)
and $\gamma=0.01 \omega_p$. 
The transition frequency is $\Omega_m/2\pi 
\approx 480\, \mathrm{MHz} 
\approx (k_B/2\pi\hbar)\, 0.023 \,\mathrm{K}$
($\lambda_m = 63 \,\mathrm{cm}$).
In both plots, the free energy scale $\mathcal{F}(1\mathrm{\mu m}) = 9.79 \times 10^{-37} \,\mathrm{J} = (2\pi\hbar) 1.48\,\mathrm{mHz}$ 
is the value at $T=0, L=1\mathrm{\mu m}$ obtained in the plasma model.
Gray dashed lines indicate the asymptotic laws at zero temperature Eqs.(\ref{eq:sub-skin-depth-regime-plasma}--\ref{eq:retarded-limit}), while colored dashed lines include thermal corrections according to Eqs.(\ref{eq:thermal-regime-Drude}, \ref{eq:non-retarded-regime-finite-T-Drude}).
}
\label{abb:l_polder_plasma&drude}
\end{figure}

%
The (free) energy vs.\ distance has been calculated numerically for
an anisotropic magnetic dipole in front of a half-space filled with a
normal conductor (Fig.\ref{abb:l_polder_plasma&drude}, top)
or described by the plasma model 
(Fig.\ref{abb:l_polder_plasma&drude}, bottom). The thick black curves
give the zero-temperature result (see the caption for parameters).
The dashed asymptotes are discussed in this section.
All energies are normalized to the $L^{-3}$ power law of the non-retarded
Casimir-Polder energy near a perfectly reflecting surface.
The scale factor $\mathcal{F}_{\rm pl}(1\mathrm{\mu m})$, given
in the caption, is slightly smaller than Eq.(\ref{eq:non-retarded-T=0})
below.
These and the following results have been obtained from the 
numerical procedure described in Appendix~\ref{a:numerics}. 

%
%

The magnetic Casimir-Polder potential is always repulsive as expected 
from the interaction between an oscillating magnetic dipole and its image at
the conducting surface. The sign is also consistent with the macroscopic
Casimir interaction between a conducting and a permeable surface (`mixed'
Dirichlet-von Neumann boundary conditions),
see e.g. Ref.\cite{Boyer74}. 
The curves in Fig.\ref{abb:l_polder_plasma&drude} make manifest
the crossovers between the distance regimes introduced in 
Sec.\ref{s:distance-regimes} above.
The relevant length scales are here the skin-depth $\delta_m = \delta_{\Omega_m}$, 
evaluated at the transition frequency (for the normal
conductor), the plasma wavelength $\lambda_{\rm p} = 
2 \pi c / \omega_p$ (for the plasma model), and the transition wavelength 
$\lambda_m = \lambda( \Omega_m )$. The case of the superconductor
is discussed in Sec.\ref{sec:superconductor} below (Fig.\ref{abb:2f-polder}):
within the two-fluid model adopted here, it is identical to the plasma model
at zero temperature. The temperature dependence interpolates between 
the Drude and plasma case, as discussed in Secs.\ref{s:plasma} and 
\ref{s:dissipative-quenching}.

The zero-temperature (black curves) case for a Drude model has been stated
earlier in Ref.\cite{Henkel05a}; we give details on the asymptotes.
Taking the limit $T \to 0$ in Eq.(\ref{eq:Matsubara-series}) recovers the well-known expression
\begin{equation}\label{FE_zero_T}
\mathcal{F}(L,0) = E(L) = - \frac{\hbar}{2 \pi} \int_0^\infty d\xi \beta_{ij} \mathcal{H}_{ji}~.
\end{equation}

In the \textit{sub-skin-depth regime} $L \ll \delta_m$, the 
distance dependence in the anisotropic case (\ref{eq:FE-aniso})
for the Drude model becomes
\begin{eqnarray}
\mathcal{F}_{\rm an}^\mathrm{Dr}(L, 0) 
&\approx& 	\frac{|\mu_x|^2 \mu_0}{8 \pi^2 \delta_m^2} \frac{1}{L} \ln\left(\frac{\delta_m}{L}\right) 
	\label{eq:sub-skin-depth-Drude}
\end{eqnarray}
where $|\mu_x|^2$ is the magnetic transition dipole matrix element, cf. Appendix \ref{Magnetic transition matrix elements}.
This expression is obtained by using 
the sub-skin-depth asymptote of the magnetic Green's tensor 
(first column of Table~\ref{MagneticAppr}) under the $\xi$-integral (\ref{FE_zero_T}) and cutting the integral off at the
border of this regime, $L \sim \delta( \xi )$, i.e., $\xi \approx 
2 \gamma c^2 / (\omega_p L)^2$. 
The small-distance calculation for the plasma model can be done in a similar
way. In both the sub-skin-depth and non-retarded regimes, the Green's 
tensor \eqref{eq:magnetic_GF} becomes independent of $\xi$
(see Table~\ref{MagneticAppr}),
and the frequency integral depends only on the polarizability. Therefore, no
logarithm appears as in the dissipative case, but
%
\begin{eqnarray}
\mathcal{F}_{\rm an}^\mathrm{pl} (L, 0) &\approx& 
\frac{|\mu_x|^2 \mu_0}{16 \lambda_{\rm p}^2} \frac{1}{L} ~.
	\label{eq:sub-skin-depth-regime-plasma}
\end{eqnarray}

In the \textit{non-retarded regime} (intermediate distances), the interaction energy in
the Drude model \cite{Henkel05a} and in the plasma model behave alike  
\begin{eqnarray}
\mathcal{F}_{\rm an}
(L, 0) &\approx& 
\frac{|\mu_x|^2 \mu_0}{32 \pi} \frac{1}{L^3} ~,
	\label{eq:non-retarded-T=0}
\end{eqnarray}
This is calculated as outlined above for the plasma model. 
The energy~(\ref{eq:non-retarded-T=0}) is identical
to the interaction of the magnetic dipole $\mu_x$ with its image,
calculated as for a perfectly conducting surface. Indeed, the $L^{-3}$
power law is consistent with the dipole field of a static (image) dipole.
%

%
In the \textit{retarded region} $L \gg \lambda_m, \delta_m, 
\lambda_{\rm p}$ (not discussed in Ref.\cite{Henkel05}),
the free energy of the Drude is identical to the one of the plasma model. Retardation effects 
lead to a change  in the power law with respect to shorter distances, 
identical to the electric Casimir-Polder interaction: 
\begin{equation}
\mathcal{F}_{\rm an}(L, 0)
= \frac{ |\mu_x|^2 \mu_0 \lambda_m }{16 \pi^3 }
\frac{ 1 }{ L^4}
~.
	\label{eq:retarded-limit}
\end{equation}
The calculation of this asymptote follows the same lines as
in the electric dipole case, see Ref.\cite{Wylie85}. Comparing
different transition wavelengths $\lambda_m$
(e.g., Zeeman vs hyperfine splitting): the smaller the transition energy,
the larger the retarded interaction.
The numerical data displayed in Fig. \ref{abb:l_polder_plasma&drude} agree very well with all three asymptotes.

Eqs.(\ref{eq:non-retarded-T=0}) and (\ref{eq:retarded-limit}) illustrate that
the magnetic atom-surface interaction is reduced relative to the
electric one by the fine-structure constant $\alpha^2_{\rm fs}$, as
anticipated earlier in Eq.(\ref{eq:magnitude}). One should bear in mind, of course, that the length
scales for the cross-overs into the retarded regime are very different.
A crossing of the non-retarded magnetic and the retarded electric
potentials would be expected for a distance of order 
$\lambda_e \alpha^{-2}_{\rm fs} \sim 1\,\mathrm{mm}$, 
 where it is clear that both energies are already extremely 
small.
In addition, the temperature should be low enough so that
the thermal wavelength ($\xi_1$ is the first Matsubara frequency)
\begin{equation}
	\Lambda_T = \frac{ c }{ 2 \xi_1 } 
	= \frac{ \hbar c }{ 4\pi k_B T } \approx \frac{0.18 {\rm mm~ K}}{T}
	\label{eq:def-thermal-wavelength}
\end{equation}
satisfies $\Lambda_T  \gg 1\,\mathrm{mm}$.
Indeed, we shall see in the following section that a nonzero temperature
can significantly reduce the magnetic Casimir-Polder potential.
\section{Casimir-Polder interaction at nonzero temperature}
\label{Casimir-Polder interaction at nonzero temperature}

In this section, we consider the temperature dependence of the 
Casimir-Polder interaction at \emph{global} equilibrium, in particular
using the temperature-dependent polarizability $\boldsymbol{\beta}^T$ (\ref{eq:two-level_polarizability}). This provides also a well-defined
calculation of the atom-surface entropy, see Sec.\ref{s:entropy}.
Scenarios with atoms prepared in specific magnetic sub-levels
are discussed in Sec.\ref{Non-thermal states}. 

The set of curves in Fig.\ref{abb:l_polder_plasma&drude} illustrates
the strong impact of a nonzero temperature for the Drude (normally
conducting) metal: its magnitude is reduced for any distance $L$.
In the plasma model (no dissipation), the main effect is the emergence 
of a different long-distance regime: the \emph{thermal regime}
$L \gg \Lambda_T$ [Eq.(\ref{eq:def-thermal-wavelength})]) 
where the interaction
becomes stronger than at $T = 0$. The latter kind of behavior could have
been expected from the thermal occupation of photon modes within 
the thermal
spectrum. The effect in the Drude model is more striking and is 
explained in Sec.\ref{s:dissipative-quenching} below.
A significant
difference with the electric dipole interaction is the fact that it is quite
common to have temperatures much larger than the magnetic resonance 
energies, $k_B T \gg \hbar \Omega_m$ or $\Lambda_T \ll \lambda_m$.
Thermal effects thus start to play a role already in the non-retarded regime,
and can be pronounced at all distances.

The usual description of the high-temperature (or Keesom \cite{Parsegian}) 
limit is based
on the term $n = 0$ in the Matsubara sum~(\ref{eq:Matsubara-series})
\begin{equation} 
\label{eq:FE_highT}
\mathcal{F}(L, T \to \infty) 
\approx - \frac{ k_B T}{2} \beta^T_{ij}(0) 
\mathcal{H}_{ji}(L,0) 
\end{equation}
Indeed, the higher terms are proportional to the small factor 
$\exp( - 2 \xi_n L / c ) = \exp( - n L / \Lambda_T )$ that appears in 
the Green's function $\mathcal{H}_{ji}(L, \imath \xi_n)$. 
This description is discussed in more detail in the following sections.

\subsection{Plasma model}
\label{s:plasma}


In the plasma model and more generally, for all materials where the
reflection coefficient $r^{\rm TE}( \omega, k )$ goes to a nonzero static
limit, the magnetic Green's tensor 
$\mathcal{H}_{ji}(L, \omega \to 0)$ is nonzero as well. 
The leading order potential
in the thermal regime is then given by Eq.(\ref{eq:FE_highT}). 
For the anisotropic polarizability of Eq.(\ref{eq:FE-aniso}), 
and assuming $k_B T \gg \hbar \Omega_m$, the temperature
dependence drops out, and 
we find from a glance at Table~\ref{MagneticAppr}
\begin{equation}
\mathcal{F}_{\rm an}^{\rm pl}(L \gg \Lambda_T, T) 
=
- |\mu_x|^2 \mathcal{H}_{xx}(L, 0) = 
\frac{ \mu_0 |\mu_x|^2 }{ 32 \pi L^3 }
	\label{eq:plasma-thermal-regime}
\end{equation}
(assuming $L \gg \lambda_p$).
This is identical to the zero-temperature result in the non-retarded
regime~(\ref{eq:non-retarded-T=0}), as can also be seen in 
Fig.\ref{abb:l_polder_plasma&drude}. If the temperature is lower,
$k_B T < \hbar \Omega_m$, but the distance still in the thermal
regime, the factor $\tanh( \hbar \Omega_m / 2 k_B T ) <
\hbar \Omega_m / 2 k_B T$ in the static polarizability
reduces the interaction slightly
($T = 0.01\,\mathrm{K}$ in Fig.\ref{abb:l_polder_plasma&drude},
bottom).


\subsection{Superconductor}
\label{sec:superconductor}

\begin{figure}[hhh]
\centering
\includegraphics[width=8cm]{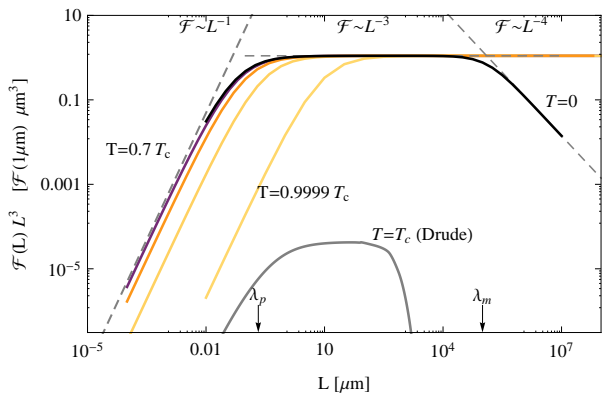}
\caption[$\mathcal{F}(L)$ anisotropic magnetic dipole, superconductor.]{(Color online) Casimir-Polder free energy vs.\ distance for an anisotropic magnetic dipole above a superconductig surface. Parameters $\omega_p$, $\gamma$, $\Omega_m$ and
normalization $\mathcal{F}(1\,\mu{\rm m})$ 
as in 
Fig.\ref{abb:l_polder_plasma&drude}, critical temperature 
$T_c =1\,{\rm K} \approx 290 \, T_m$. Temperatures are $T/T_c \in \{0, 0.7, 0.9, 0.99,0.9999, 1.0\}$.
%
}
\label{abb:2f-polder}
\end{figure}

The atom-superconductor interaction shows a richer behavior compared
to the plasma model, as illustrated in Fig.\ref{abb:2f-polder}.
%
%
At $T = 0$, it strictly coincides with the plasma model, as it must 
for the two-fluid description~(\ref{eq:epsilon-two-fluid-model}) adopted here. 
The large-distance (thermal) asymptotes are the same as in the plasma
model for $T < T_c$. The reasoning leading to 
Eq.(\ref{eq:plasma-thermal-regime}) can be applied here as well:
the response of the superconducting surface to a static magnetic field
is characterized by a nonzero value for $r^{\rm TE}(\omega \to 0, k)$
because of the Mei\ss{}ner-Ochsenfeld effect. Although the superconducting fraction 
decreases to zero, proportional to the product $\eta(T)\omega_p^2$,
the interaction potential Eq.(\ref{eq:plasma-thermal-regime}) stays
constant because it does not depend on this `effective plasma frequency'.

This picture also explains the lowering of the sub-skin-depth
asymptotes in Fig.\ref{abb:2f-polder}: from 
Eq.(\ref{eq:sub-skin-depth-regime-plasma}), the Casimir-Polder potential
is proportional to $1/\lambda_p^2 \mapsto \eta(T)(\omega_p/c)^2$.
This gives scale factors $\approx \frac12, \ldots, 2\times 10^{-4}$ for 
the cases
$T = 0.7, \ldots, 0.9999\,T_c$, in quite good agreement with the 
numerical data.

\begin{figure}[hhh]
\includegraphics[width=8cm]{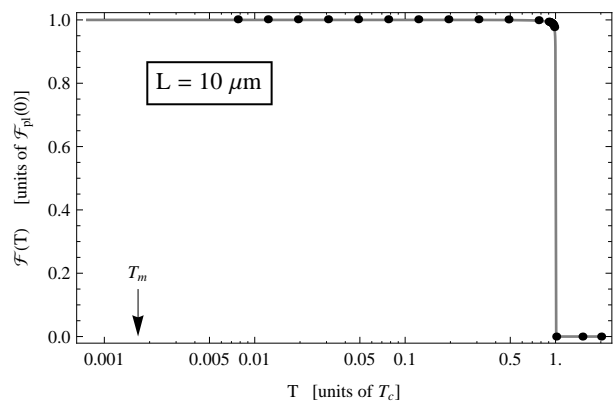}
\caption[$\mathcal{F}(T)$ anisotropic magnetic dipole, superconductor.]{
Casimir-Polder 
free energy for a superconducting surface, in the
two-fluid model (solid line) and the BCS theory (dots).
Parameters $\omega_p$, $\Omega_m$ as in 
Fig.\ref{abb:l_polder_plasma&drude}. 
Scattering rate $\gamma = 5\times 10^{-4} \omega_p$,
critical temperature $T_c = 12\,\mathrm{K} = 500 \, T_m$,
atom-surface distance $L = 1\,\mathrm{\mu m}$. 
Energies normalized to $\mathcal{F}_{\rm pl}(L, T \to 0) 
= 1.09\times10^{-39} \mathrm{J}$.
}
\label{abb:mag_polder_BCS}
\end{figure}

It is worth mentioning that the full BCS theory can give results in close 
agreement with the simple two-fluid model we use here. In 
Fig.\ref{abb:mag_polder_BCS}, we show the temperature dependence
of the Casimir-Polder potential (at fixed distance $L$) for the two cases.
We choose here a
damping parameter $\gamma = 5\times 10^{-4} \omega_p$ in the same order of the zero-temperature gap $ \Delta(0) = 3.5\times 10^{-4}\hbar \omega_p \approx 1.76 k_B T_c$.

The BCS calculations have been performed using a recently developed
efficient technique of calculating the optical conductivity at imaginary 
frequencies \cite{BCSunpub} and using an approximative form of the gap equation \cite{Townsend1962, Thouless1960}.
Calculations over a larger parameter range, but restricted to
$T = 0$, have been reported by Skagerstam \emph{et al.}
\cite{Skagerstam09}.


Going back to Fig.\ref{abb:2f-polder}, note that
at $T > T_c$, the superconductor jumps to a completely different
behavior, identical to the Drude metal. This is expected from the
two-fluid model~(\ref{eq:epsilon-two-fluid-model}), but also in
Mattis-Bardeen theory where the gap parameter $\Delta( T )$ vanishes
above $T_c$, and the
optical conductivity $\sigma( \omega, T )$ coincides with the Drude
model, see Refs.\cite{Mattis58,Zimmermann91,Berlinsky93}.

\subsection{Thermal decoupling from a normal conductor}
	\label{s:dissipative-quenching}
	\label{Dissipative quenching}

As mentioned above, the Drude model and the superconductor around
the critical temperature show an unusually strong temperature dependence
in the magnetic Casimir--Polder potential. The strong suppression at large
distances (Figs.\ref{abb:l_polder_plasma&drude} and \ref{abb:2f-polder})
arises from the fact that the Green's tensor $\mathcal{H}_{ij}( L, {\rm i}\xi )
\to 0$ at zero frequency in the normal conducting state. 
The leading order potential~(\ref{eq:FE_highT}) vanishes, and
one has to consider the next term $\xi = \xi_1$ in the
Matsubara sum~\ref{eq:Matsubara-series}, so that the 
exponentially small factor $\exp( - L / \Lambda_T )$ governs
the thermal (large distance) regime.
We call this the \emph{thermal decoupling} of the
atom from the (normal) metal. This phenomenon is related to low-frequency
magnetic fields that penetrate the (non-magnetic) surface. Indeed, the
vanishing of $\mathcal{H}_{ij}( L, \omega \to 0 )$
could have been expected from the Bohr-van-Leeuwen 
theorem~\cite{Leeuwen21, Bimonte09} that states that for any classical
system, the magnetization response to static fields must vanish. Both
conditions apply here: the zeroth term in the Matsubara series 
involves static fields, and is also known as the classical limit. Indeed, 
except for the material coupling
constants, Eq.(\ref{eq:FE_highT}) no longer involves $\hbar$, while
the next Matsubara terms do (via $\Lambda_T$). The Bohr-van-Leeuwen
theorem does not apply to a superconductor whose response is a quantum 
effect (illustrated, for example, by the macroscopic
wave function of Ginzburg-Landau theory), and by extension, 
not to the plasma model,
as recently discussed by Bimonte \cite{Bimonte09}.

We now calculate the next order in the Matsubara series to understand
the temperature dependence of the Casimir-Polder shift near a metal.
For simplicity, we consider again the limiting case $k_B T \gg \hbar 
\Omega_m$ which simplifies the polarizability to the Keesom form,
\begin{equation}
	\beta_{\rm an}( \imath \xi_n ) \approx 
\frac{ |\mu_x|^2 \Omega_m^2 }{ k_B T \xi_n^2 }~,
\qquad n \ge 1
.
	\label{eq:Keesom-polarisability}
\end{equation}
In the thermal regime, $L \gg \Lambda_T$, we use the 
large-distance limit of the Green's tensor (cf.\ the retarded regime of 
Table~\ref{MagneticAppr}). The Matsubara frequency $\omega = 
{\rm i}\xi_1$ then yields the mentioned exponential suppression
\begin{equation}
	\mathcal{F}^{\rm Dr}_{\rm an}( L, T ) \approx
	\frac{ \pi \mu_0 |\mu_x|^2 }{ \lambda_m^2 L }
	\exp( - L / \Lambda_T )
	\label{eq:thermal-regime-Drude}
\end{equation}
where $\lambda_m$ is the magnetic resonance wavelength (cf. Fig.\ref{abb:l_polder_plasma&drude}).

At shorter distances, we have to perform
the Matsubara summation. In the regime $L \ll \Lambda_T \ll
\lambda_m$, we consider the non-retarded approximation to
the Green's tensor and make the approximation $\exp( - n L / \Lambda_T )
\approx 1$.  The sum over the polarizability
$\beta_{\rm an}( \imath \xi_n )$ can then be done with the
approximation~(\ref{eq:Keesom-polarisability}), and we get
\begin{equation}
	\mathcal{F}^{\rm Dr}_{\rm an}( L, T ) \approx
	\frac{ \mu_0 |\mu_x|^2 }{384 \pi L^3 }
	\left( \frac{ \hbar \Omega_m }{ k_B T }
	\right)^2
	\label{eq:non-retarded-regime-finite-T-Drude}
\end{equation}
The scaling $T^{-2}$ is in good agreement with Fig.\ref{abb:l_polder_plasma&drude}. The crossover into the sub-skin-depth
regime is now temperature-dependent and occurs where the 
skin depth $\delta( \xi_1 ) \sim L$. This corresponds to a
temperature $k_B T_D \sim \hbar \gamma \lambda_{\rm p}^2 / (2 \pi L)^2$.
The involved frequency $\xi_L$ is characteristic for the diffusive transport 
of electromagnetic radiation in the metal at wavevectors $\sim 1/L$.

In the sub-skin-depth regime, the leading order approximation to the
Matsubara sum involves terms up to a frequency $\xi_n \sim \xi_L$.
This leads to an asymptote similar to Eq.(\ref{eq:sub-skin-depth-Drude}),
but with the ratio $\hbar \xi_L / k_B T$ in the argument of the
logarithm and an additional factor $\hbar \Omega_m / k_B T$.

\section{Atom-surface entropy}
\label{s:entropy}

It is well-known that at high temperatures where the free energy
scales linearly in $T$ [Eq.(\ref{eq:FE_highT})], the dispersion interaction
is mainly of entropic origin \cite{Parsegian}. 
More precisely, the interaction is proportional to the change in entropy of 
the system 
``atom plus field plus 
metallic surface'',
as the atom is brought from infinity to a distance $L$ from the surface. 
We calculate in this section the atom-surface entropy according to
\begin{equation}
	S( L, T ) = - \frac{ \partial \mathcal{F} }{ \partial T }
	\label{eq:def-entropy}
\end{equation}
This entropy definition is unambiguous for the global equilibrium setting
of the previous Section. 

The behavior seen in the previous figures indicates significantly
different entropies for the surface models, with a strong dependence on 
the presence of dissipation (conductivity) at low frequencies.
This parallels
the discussion of the macroscopic Casimir entropy for the dispersion
interaction between two plates, a subject of recent controversies,
where the Drude and plasma models give different answers
\cite{Bezerra04,Bostrom04,Svetovoy05,Sernelius06,Intravaia08,Klimchitskaya09}. The results
that follow indicate that the magnetic Casimir-Polder interaction may
provide an alternative scenario to investigate this point. 


\begin{figure}[hhh]
\centering
\includegraphics[width=8cm]{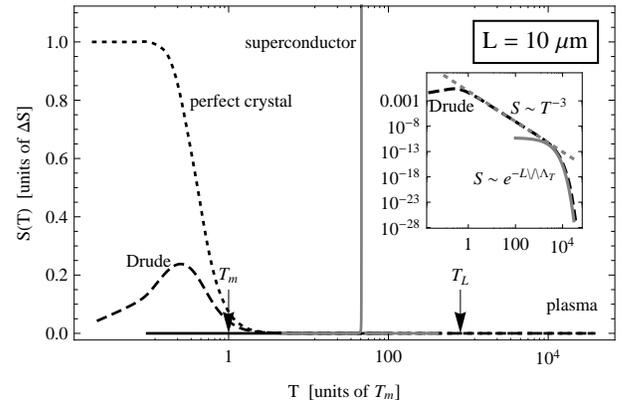}
\caption[$\mathcal{F}(T)$ anisotropic magnetic dipole.]{
 Casimir-Polder 
entropy of an anisotropic magnetic dipole with a plasma (solid), Drude-metal (dashed), perfect crystal (dotted) and two fluid superconductor (gray). 
Parameters chosen as in Figs.\ref{abb:l_polder_plasma&drude} and
\ref{abb:2f-polder}.
The unit of entropy $\Delta S$ is defined in Eq.\eqref{eq:delta_entropy}.
The inset shows a the data for the Drude model and its asymptotics obtained from Eqs.(\ref{eq:thermal-regime-Drude},\ref{eq:non-retarded-regime-finite-T-Drude}) in logarithmic scaling.
}
\label{abb:mag_polder}
\end{figure}


The atom-surface entropy~(\ref{eq:def-entropy}) is plotted in 
Fig.\ref{abb:mag_polder} for surfaces of different material. 
In all models, the
entropy vanishes at high temperatures because to leading order, the
free energy~(\ref{eq:FE_highT}) becomes independent of $T$,
and higher orders vanish exponentially with $T$. (This
feature is specific to the thermal polarizability of a two-level system.)
It is remarkable that the vanishing of the entropy happens at temperatures
already much smaller than the `geometrical scale' $k_B T_L 
= \hbar c / 4\pi L$ [i.e. $T /  T_L = 2 \xi_1 L / c$ with $\xi_1$
the first
Matsubara frequency]. This points towards another characteristic
energy scale in the atom-surface system, discussed below.


One notes in Fig.\ref{abb:mag_polder}
very small values for a plasma and a superconductor,
two cases where the dc conductivity diverges. 
The superconductor shows a narrow, pronounced entropy peak at $T_c$:
we interpret this as the participation of the atomic dipole in the phase
transition. Indeed, the electromagnetic waves near the surface are slightly 
shifted in phase due to the interaction with the magnetic dipole moment. The 
atom-surface interaction can be thought of a sum over all these phase
shifts, similar to Feynman's interpretation of the Lamb shift.


In the Drude model, we observe a broad peak at temperatures where
the thermal energy $k_B T$ becomes comparable to the photon energy
of the magnetic resonance, $\hbar \Omega_m \equiv k_B T_m$. 
Comparable to this scale for our parameters is the diffusive energy
$k_B T_D \sim \hbar \gamma \lambda_{\rm p} / (2\pi L)^2 \approx
3.3 \, \hbar \Omega_m$, introduced after
Eq.(\ref{eq:non-retarded-regime-finite-T-Drude}). We thus attribute
the atom-surface entropy to the participation of the atom in the thermally
activated diffusive motion of charges and fields below the metal surface
(Johnson-Nyquist noise).
This motion involves, at the relevant low energies, mainly eddy currents
whose contribution to the Casimir entropy (in the plate-plate geometry)
has been recently discussed by two of us~\cite{Intravaia09}. As $T$
drops below the diffusive scale $T_D$, the eddy currents `freeze' to
their ground state and the entropy vanishes linearly in $T$.


The dotted curve in Fig.\ref{abb:mag_polder} corresponds to the
`perfect crystal' that has not been discussed so far. It gives rise
to a nonzero atom-surface entropy in the limit $T \to 0$ which
is an apparent violation of the Nernst heat theorem (third law of
thermodynamics). This has also been discussed for the two-plate
interaction~\cite{Bezerra04,Intravaia08,Ellingsen09b}, but the entropy 
defect here has a different sign (it is negative for two plates). The sign 
can be attributed to our atomic polarizability being paramagnetic, while
metallic plates show a diamagnetic response. Using the technique
exposed in Ref.\cite{Intravaia08}, the limit of the atom-surface entropy 
as $T \to 0$ can be calculated, with the result for an anisotropic 
dipole:
\begin{eqnarray}
\frac{ \Delta S( L ) }{ k_B }
	&=& 
- \beta^0_{\rm an}(0) \mathcal{H}_{xx}^{\rm pl}(L, \omega \to 0)
\nonumber\\
   & \approx &
\frac{ \mu_0 |\mu_x|^2 }{ 16 \pi \, \hbar \Omega_m }
\frac{ 1 }{ L^3 }
~.
	\label{eq:delta_entropy}
\end{eqnarray}
The second line applies in the non-retarded limit $L \gg \lambda_{\rm p}$.
This expression is used to normalize the data in Fig.\ref{abb:mag_polder}
and provides good agreement for $T \gg T_m$. 

One can argue along the same lines as in Ref.\cite{Intravaia09} that
the Nernst theorem is actually not applicable for this system, since 
the perfect crystal cannot reach equilibrium over any finite time
in the limit of vanishing dissipation, $\gamma \to 0$. The 
entropy $\Delta S$ then describes the modification that the atom imposes
on the ensemble of field configurations that are `frozen' in the perfectly
conducting material.

In the two-plate scenario considered in Refs.\cite{Svetovoy05, Sernelius06} it has been shown that not only dissipation but also nonlocality of the response has strong implications for the entropy. In particular, the residual entropy $\Delta S=0$ vanishes, because at very low temperatures, the anomalous skin effect and Landau damping take the role of a nonzero dissipation rate.
Though we have not considered nonlocality in this work explicitly, one can expect  the same thing to happen in the magnetic Casimir-Polder interaction.

\begin{figure}[hhh]
\includegraphics[width=8cm]{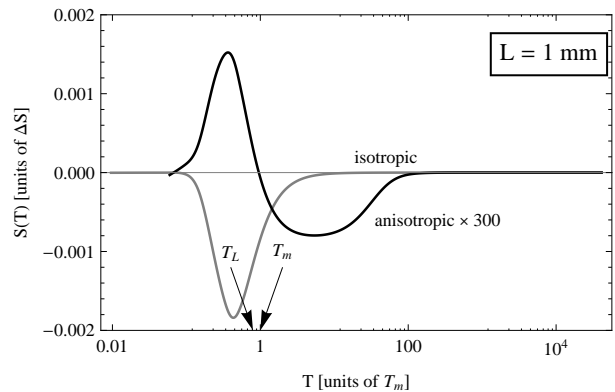}
\caption[$S(T)$ anisotropic and isotropic magnetic dipole, plasma.]{
Casimir-Polder 
entropy of a magnetic dipole with a surface described in the plasma model, 
anisotropic and isotropic polarizability.
Parameters $\omega_p$, $\Omega_m$ as in Fig.\ref{abb:l_polder_plasma&drude}, 
atom-surface distance
$L=1\,\mathrm{mm}$ where the thermal effects are better visible. 
Same entropy scale factor $\Delta S$ as in 
Fig.\ref{abb:mag_polder}, defined in Eq.(\ref{eq:delta_entropy}).
The anisotropic curve is blown up to become visible.
}
\label{abb:mag_polder_plasma}

\end{figure}

It should be mentioned, that we have also found negative values for the 
atom-surface entropy, albeit very small, for temperatures between $T_m$
and $T_L$ and in the retarded regime
(see Fig.\ref{abb:mag_polder_plasma}). The sign depends on the orientation
of the dipole and is sensitive to a balance between the TE- and TM-polarized
parts of the magnetic Green's tensor.


\section{Non-thermal states}
\label{Non-thermal states}

We have argued in the last section, that many realistic setups involve 
non-equilibrium situations. Atom-chips are a typical example where 
two independent phenomenological temperatures can be introduced for the 
atom (or a sample of atoms) and the surface. This temperature gradient
is metastable on experimentally relevant time scales because of the weak
interaction between the subsystems.

We now analyze the case where the surface is described by a temperature
$T$, and the atom prepared in a well-defined state $|a\rangle$. 
More complex configurations 
can be studied starting from this simple case.
We first consider two-level atoms 
and then multilevel atoms, including hyperfine transitions as they occur for
the alkali group.

\subsection{Two-state atom}

As before, there is a single resonance frequency $\Omega_m$ for
the two-level atom. Depending on whether the atom is prepared in 
the ground-state $| g \rangle$ or the excited state $|e\rangle$,
the sign of the polarizability~(\ref{eq:WS_polarizability}) changes.
Referring to the first line of Eq.(\ref{WS}), $\boldsymbol{\beta}^a$ does no longer
depend on temperature. The resonant term of the second line
involves the thermal occupation number that we approximate by its
classical value
$n( \pm \Omega_m )\approx \pm k_{B}T/(\hbar\Omega_m)$.
This is sufficiently accurate at room temperature and typical magnetic
resonances. 
In this limit, we obtain a simple expression for the magnetic
Casimir-Polder free energy (anisotropic case, argument `$g$' for
ground-state atom)
\begin{eqnarray}
\label{WS2Level}
&&
\mathcal{F}_{\rm an}(L, g, T)
\approx
- 
2 k_{B}T \sum_{n\ge 1}\beta^{a}(\imath\xi_{n})
\mathcal{H}_{xx}(L,\imath\xi_{n})
\\
&& {} \qquad + k_{B}T\beta^{a}(0)\left\{
\re[\mathcal{H}_{xx}(L,\Omega_{m})]
-
\mathcal{H}_{xx}(L,0)
\right\}
\nonumber~.
\end{eqnarray}
The first line is similar to the result in the Drude model because 
of the missing zeroth Matsubara term. From this expression, we
now discuss the differences between the Drude and the plasma model.

\begin{figure}[h]
\centering
\includegraphics[width=8cm]{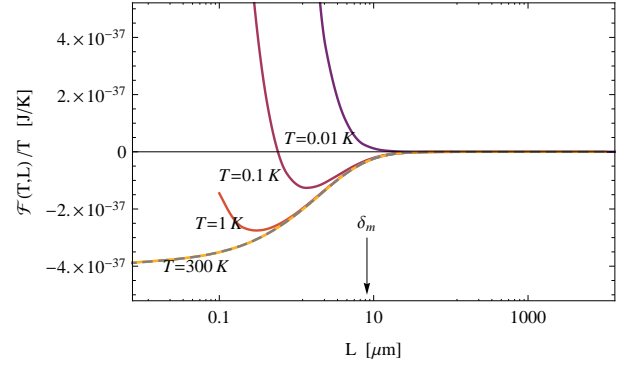}
\caption[$\mathcal{F}(L)$ ground state atom.]{(Color online) %
Casimir-Polder free energy vs.\ distance for a two-level atom in its ground 
state, transition dipole parallel to the surface. 
The surface is described by a Drude metal at different temperatures.
Parameters $\omega_p$, $\gamma$, $\Omega_m$ as in 
Fig.\ref{abb:l_polder_plasma&drude}. Note the scale factor $1/T$
to show the classical limit $\mathcal{ F } \sim T$.
The limiting curve at high temperatures (gray dashed line)
can be inferred from the second line of Eq.(\ref{WS2Level}).
}
\label{abb:ooe_mag_polder}
\end{figure}

\begin{figure}
\includegraphics[width=8cm]{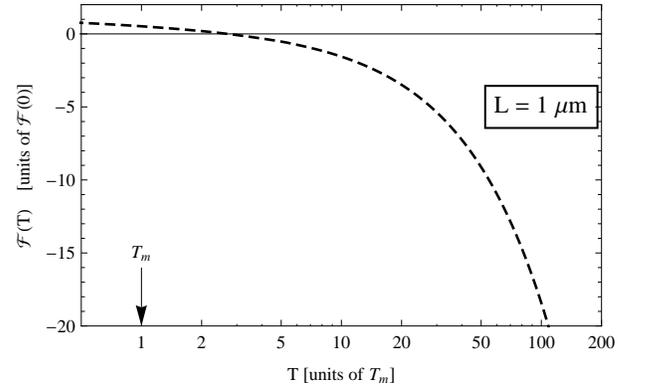}
\caption{%
Same as Fig.\ref{abb:ooe_mag_polder}, but vs.\ temperature. 
Distance $L = 1\,\mu\mathrm{m}$.
The energy scale $\mathcal{F}(0) =2.56 \times 10^{-38} \mathrm{J} $ 
is the value at $T=0\mathrm{K}$.
}
\label{abb:ooe_mag_polder_2}

\end{figure}

For the Drude model, the resonant contribution involving 
$\re{ \mathcal{H}(L,\Omega_{m})}$ becomes significant in the
non-retarded regime.
In particular, combined with the non-resonant contribution, 
it changes the sign of the Casimir-Polder potential already at short
distances,  resulting
in an \emph{attractive interaction}, as soon as $T \gtrsim T_m$,
see Figs.\ref{abb:ooe_mag_polder} and \ref{abb:ooe_mag_polder_2}.




\begin{figure}[h]
\centering

\includegraphics[width=8cm]{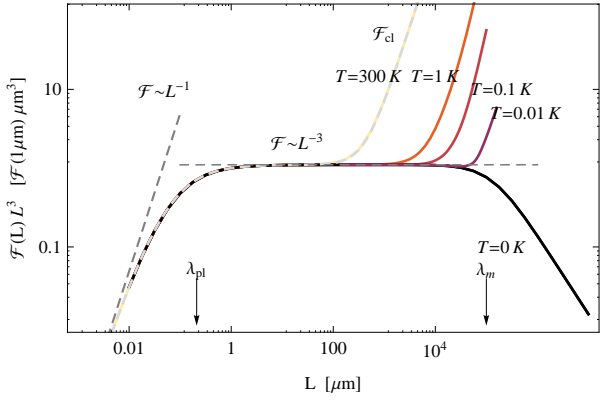}
\caption[$\mathcal{F}(L)$ ground state atom, plasma.]{(Color online) 
Casimir-Polder free energy vs.\  distance for a two-level atom
near a plasma surface. 
Atom in the ground state and transition dipole parallel to the surface.
Parameters $\omega_p$, $\Omega_m$ and scale factor
$\mathcal{F}(1\mathrm{\mu m})$ as in 
Fig.\ref{abb:l_polder_plasma&drude}(bottom). At high temperatures, the curve can be well approximated by the classical contribution (thick gray dashed line). The oscillating part of Eq.(\ref{eq:retarded-n-eq-plasma}) not shown here, sets in at distances $L \gtrsim \lambda_m$.
}
\label{abb:ooe_mag_polder_plasma}

\end{figure}

In contrast, the second line of Eq.(\ref{WS2Level}) nearly vanishes 
in the plasma model because the 
magnetic Green's function $\mathcal{H}_{xx}(L, \omega)$ is 
approximately independent of frequency,
at least in the non-retarded regime. 
We thus get a situation
where the zeroth Matsubara term is nearly removed from the Casimir-Polder
potential. The resonant term still gives the leading order contribution,
once the expansion of the occupation number is pushed to 
the next order,
$n( \Omega_m ) \approx k_B T / \hbar \Omega_m - \frac{ 1}{2 }$. 
We then get
\begin{equation}
	\mathcal{F}^{\rm pl}_{\rm an}( L, g, T ) \approx
	- |\mu_x|^2 \re \mathcal{H}_{xx}( L, \Omega_m ) 
	\approx
	\frac{ \mu_0 |\mu_x|^2 }{ 32\pi L^3 }
	\label{eq:non-retarded-n-eq-plasma}
\end{equation}
where the last expression applies in the non-retarded regime
and is identical to the $T = 0$ case [Eq.(\ref{eq:non-retarded-T=0})], cf. Fig.\ref{abb:ooe_mag_polder_plasma}.
At larger distances (retarded regime), the difference between the 
Green's functions in the second line of Eq.(\ref{WS2Level}) is nonzero 
and becomes the leading term:
\begin{eqnarray}
	\mathcal{F}^{\rm pl}_{\rm an}( L, g, T ) &\approx&
	\frac{ k_B T }{ \hbar \Omega_m }
	\frac{ \mu_0 \pi |\mu_x|^2 }{ \lambda_m^2\, L}
	\nonumber\\
&& 
	\times \left[
	\cos\frac{ 4 \pi L }{ \lambda_m}
	- \frac{ \lambda_m }{ 4 \pi L }
	\sin\frac{ 4 \pi L }{ \lambda_m} 
	\right]
	\label{eq:retarded-n-eq-plasma}
\end{eqnarray}
Note that this has a longer range than the $1/L^3$ power 
law~(\ref{eq:non-retarded-n-eq-plasma}), see Fig.\ref{abb:ooe_mag_polder_plasma}.
This effect is well known from the electric-dipole interaction of
excited atoms \cite{Hinds91} and consistent with the classical interpretation
(frequency shift of an antenna) of the resonant term.

\subsection{Trapped rubidium atom}
\label{Trapped rubidium atom}

The atom-surface potential now involves transitions 
$|a\rangle \leftrightarrow |b\rangle$ to both higher and lower energy 
levels. Eq.(\ref{WS}) yields the following form of the free energy
\begin{align}
\label{WSGeneral}
& \mathcal{F}(L, a, T) \approx
- k_{B}T \sum_{n\ge 1, j}\beta^{a}_{jj}(\imath\xi_{n})
\mathcal{H}_{jj}(L,\imath\xi_{n})
\\
& {} \qquad + k_{B}T
\sum_{b,j} 
\frac{|\mu_j^{ab}|^{2}}{\hbar\omega_{ba}} 
\left\{
\re [\mathcal{H}_{jj}(L,\omega_{ba})]
-
\mathcal{H}_{jj}(L,0)
\right\}~,
\nonumber
\end{align}
where we assume again 
that $k_B T\gg \hbar|\omega_{ba}|$ which is valid in many experiments. 
The sign of the interaction depends on the relative weight of virtual
transitions to lower and higher energy levels. From 
Eq.(\ref{eq:WS_polarizability}), a virtual level  $E_b > E_a$ gives a
positive contribution to the polarizability and a positive prefactor for the 
second line in Eq.(\ref{WSGeneral}), these terms being negative for
levels $E_b < E_a$. We can interpret this sign change from the 
difference between stimulated emission into the thermal radiation field
(for the excited atom) and photon absorption (for the ground state).
Generally speaking, these contributions do not cancel each other
because the matrix elements $|\mu_j^{ab}|^2$ are not the same.

\begin{figure}[hhhh]
\includegraphics[width=5cm]{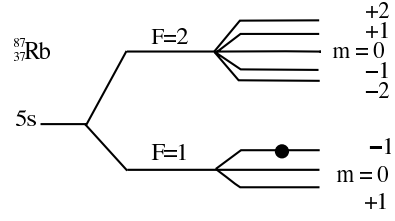}
\caption{Energy scheme for $^{87}\rm Rb$.}
\label{fig:rb87}
\end{figure}

Let us consider the example of $^{87}\mathrm{Rb}$, prepared
in the magnetically trappable hyperfine state $|a\rangle = | F, m_F \rangle
= |1, -1\rangle$ of the $5s$ ground state configuration (Fig.\ref{fig:rb87}).  This atom has
vanishing orbital momentum $L = 0$, nuclear spin $I = 3/2$, and 
a single valence electron so that $J = S = 1/2$. 
The splitting between the hyperfine levels $F=1$ (lower)
and $F = 2$ is $\Omega_{\rm hf} / 2 \pi \approx 6.8 \,
\mathrm{GHz}$, to which the Zeeman splitting in the magnetic trap
must be added with the correct Land\'e factor.
We use the same Larmor frequency as before for the
two-level atom; because of
$\Omega_m / \Omega_{\rm hf} \approx 0.07$,
we are still in the weak-field regime.

\begin{figure}[hhh]
\centering
\includegraphics[width=8cm]{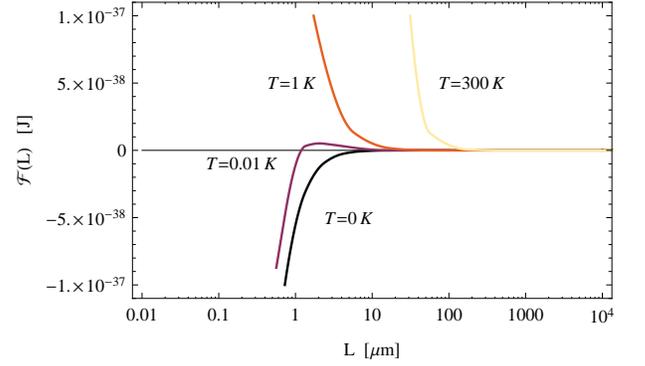}
\caption[$\mathcal{F}(T)$, excited atom, Drude.]{(Color online) Magnetic Casimir-Polder free energy near a Drude metal 
for a $^{87}\mathrm{Rb}$ atom in a given hyperfine state
($|F, m_F\rangle = |1,-1\rangle$). 
Parameters $\omega_p$, $\gamma$ as in Fig.\ref{abb:l_polder_plasma&drude}.
The Larmor frequency (for virtual transitions between neighboring Zeeman levels)
has the same value $\Omega_m/2\pi = (k_B/2\pi\hbar)\, 23\,\mathrm{mK}$ as before,
and the hyperfine splitting is
$\Omega_{\rm hf} / 2 \pi \approx 6.8 \, \mathrm{GHz}
\approx (k_B/2\pi\hbar)\, 0.3\,\mathrm{K}$.
}
\label{abb:ooe_mag_RB_Drude}
\end{figure}

\begin{figure}[tbh]
\centering
\includegraphics[width=8cm]{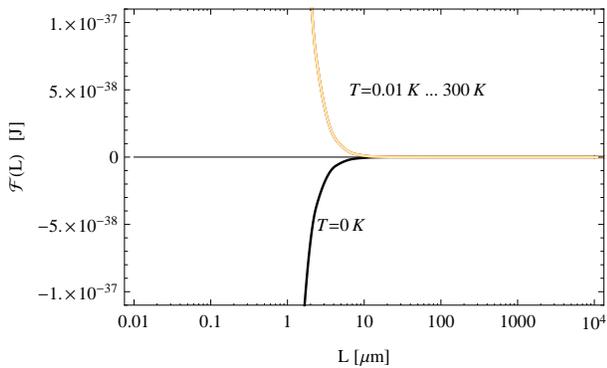}
\caption[$\mathcal{F}(L)$ excited atom, plasma.]{(Color online) Same as Fig.\ref{abb:ooe_mag_RB_Drude}, but for the
plasma model.
The
numerical calculations for extremely low temperatures $0 < T \ll
T_m$, where the potential changes sign,
could not be performed with sufficient precision.  
}
\label{abb:ooe_mag_RB_Plasma}

\end{figure}

Assuming a quantization axis perpendicular to the surface, see 
Sec.\ref{s:optical-magnetic-traps}, we get an anisotropic 
polarizability. The necessary matrix elements are calculated in 
Appendix~\ref{Magnetic transition matrix elements}.
Numerical results for the Casimir-Polder interaction 
according to Eqs.\eqref{WS} and \eqref{eq:WS_polarizability},
are shown in Figs.~\ref{abb:ooe_mag_RB_Drude} and
\ref{abb:ooe_mag_RB_Plasma} for the Drude and plasma models,
respectively.

Near a normal conducting surface described by the Drude model, 
the interaction for $T = 0$ is attractive at all distances. 
We associate
the sign reversal (compared to the absolute ground state considered 
so far) to the coupling to the lower-lying
Zeeman levels.
At high temperatures, the interaction becomes dominated by
the resonant contribution that grows linearly with $T$ and
is repulsive. Again, we find
the opposite sign as for the ground-state atom in 
Fig.\ref{abb:ooe_mag_polder}.  Thus, the sum of both
contributions leads to a maximum of the free energy
at a nonzero, $T$-dependent distance.

In the plasma model, the potential crosses over globally (for
all distances) from attractive to repulsive. For all practical
temperatures, $T > T_m$, and the interaction will be repulsive
as shown in Fig.\ref{abb:ooe_mag_RB_Plasma}.
The results illustrate that the magnetic dipole interaction of an excited
atom will be repulsive in all practical realizations.  

To summarize, the strong dependence of the thermal correction
on dissipation in the surface occurs in both 
non-equilibrium situations considered here: two-level or multilevel
atoms prepared in a given energy state. The magnetic Casimir-Polder
potential thus offers an
opportunity to distinguish between the two models on the basis of
experimental data taken at low surface temperatures and small
distances within the possibilities of today's experiments.

\section{Conclusion and Discussion}
\label{Conclusion and Discussion}
We have considered the interaction of a magnetically polarizable
particle with a metallic or superconducting surface including the
effects of nonzero temperature and out-of-equilibrium situations.
Previous work had considered mostly the case of electric
polarizability, e.g., Ref.\cite{Bezerra08}, or was limited to a static
magnetic dipole \cite{Bimonte2009a} or zero temperature
\cite{Henkel05a, Skagerstam09}.
 
The magnetic atom-surface interaction is repulsive over a large range of parameters and turns out to be highly sensitive
to both thermal fluctuations and dissipation.
In this respect, it shows similarities
with the Casimir interaction between metallic or magneto-dielectric plates.

The $T=0$ results of Ref.\cite{Henkel05a} suggested that the
magnetic interaction might be enhanced by raising the temperature,
possibly creating a regime where it dominates over the electric
contribution. In fact, thermal enhancement occurs only near a superconductor
at distances beyond the thermal wavelength, where the material response is governed by the Meissner effect.
In normal conductors, field penetration prevents such a regime -- in accordance with the Bohr-Van Leeuwen theorem -- and the Casimir-Polder energy is exponentially suppressed in global equilibrium.

This behavior can be understood qualitatively from the competition between attractive and repulsive contributions to the force.
Repulsion arises from the fluctuations of the magnetic dipole, coupled to its mirror image. This is similar to the interaction between electric currents (Lenz rule).
Field fluctuations, on the other hand, produce attractive forces, due to the paramagnetic character of the atom polarizability.
Both contributions differ in their temperature dependence and depend on the state of the atom (thermalized, spin polarized). For example, attractive forces arise between a ground-state atom and a normal conductor, as the temperature scale exceeds the magnetic transition energy. Under realistic conditions, this flips the sign of the interaction in the regime accessible to experiments (Fig.\ref{abb:ooe_mag_polder}).

Considering the Casimir-Polder entropy, we found that atoms probe the fluctuations in the material: for instance, at the superconducting phase transition a pronounced peak appears.
In most situations, the entropy vanishes at absolute zero in agreement with Nernst's theorem. The only exception we found was the particular case of a `perfect crystal' [Drude dissipation rate  $\gamma(T)\propto T^{n},\ n>1$], already discussed thoroughly in the context of the two-plate Casimir interaction
\cite{Milton09a,Klimchitskaya09}. Here, the entropy takes a (positive) nonzero value at 
zero temperature and can be understood as the participation of the atom in the disorder entropy of currents frozen below the surface \cite{Intravaia09}.
Indeed, the magnetic dipole moment mimics the current response of a second plate, except for the sign.

If we compare with an electric dipole, the temperature dependence of the Casimir-Polder interaction is relatively weak there. This is due to the larger value of the transition frequency, so that for realistic temperatures there is no difference between the ground-state and the thermalized polarizability. This was assumed in Ref.\cite{Bezerra08}, where it was also stated that all good conductors behave very similarly. In fact, the electric dipole coupling is dominated by TM-,polarized modes which cannot penetrate into the bulk due to screening by surface charges.  

We have shown here that the Casimir-Polder interaction between a metal and a fluctuating magnetic dipole resembles in many respects the Casimir interaction between two metallic plates.
In both scenarios, the thermal dependence is much more pronounced for a 
normally conducting surface compared to a description without Ohmic losses.
The role of dc conductivity in the two-plate scenario is still an open question at nonzero temperature. The atom-surface interaction may thus provide an alternative way to investigate the temperature dependence of the Casimir effect, e.g. in atom chip experiments.
The main challenge is the small value of the
interaction energy as compared to the electric one.
Future measurements of the magnetic Casimir-Polder interaction may involve high precision spectroscopy on the shift in
hyperfine or Zeeman levels.
In order to separate effects of the magnetic and
the electric dipole coupling it will be important to find control parameters that affect exclusively the magnetic contribution.
This may include the variation in external fields, isotope shifts, and highly polarizable atomic states like Rydberg atoms.

\section*{Acknowledgments}
The work of H.H. and C.H. was supported by grants from the 
 German-Israeli Foundation for Scientific Research and Development (GIF).
F.I. thanks the Alexander von Humboldt foundation for financial support. The Potsdam group also received support from the European Science Foundation (ESF) within the activity `New Trends and Applications of the Casimir Effect'. 
F.S. acknowledges support from MICINN (Spain), through Grant No. FIS-2007-65723, and the Ramon Areces Foundation.
R.P. and S.S. acknowledge partial financial support from Ministero dell'Universit\`a  e della Ricerca Scientifica and from Comitato Regionale di Ricerche Nucleari e di Struttura della Materia.
 
\appendix
\section{Numerics}
\label{a:numerics}

The thermal free energies have then been calculated numerically in the
Matsubara-formalism by summing over a thermal spectral free energy
density $g(i \xi) = -k_B T \beta_{ij} \mathcal{H}_{ij}(i\xi,L)$ at discrete imaginary frequencies.  Now, the infinite sum is
replaced by a finite one plus an integral to approximate the remaining
infinite partial sum
\begin{eqnarray}
\label{eq:Euler_MacLaurin_Num}
\mathcal{F}(T)= {\sideset{}{'}\sum_{n=0}^{\infty}}
g(i \xi_n) \approx {\sideset{}{'}\sum_{n=0}^{N}} g(i \xi_n) + \int_N^\infty g(i \xi_n) dn~.~
\end{eqnarray}
If the upper summation limit $N$ is chosen correctly, the remainder is a slowly varying function and the error is small, according to the Euler-MacLaurin formula.
In all systems considered in this work, the exponential in the Green's function \eqref{eq:magnetic_GF} ensures this property.

Numerical calculations require an automatic and fast estimate for the summation limit $N$.
In a similar scheme a fixed number of Matsubara terms ($N=10$) has recently been proposed \cite{Bostrom04}.
It should be pointed out, that since the Matsubara frequencies are linear in $T$, any such scheme with a fixed number of terms breaks down at low temperatures, where the integrand $g(i \xi)$ has not sufficiently decayed.
A basic criterion for $N$ is that the partial sum is sufficiently large, so that the integral  is only a small correction
\begin{eqnarray}\label{eq:n_numerik}
\frac{\int_{N}^{\infty} g(i \xi_n) dn }{{\sum_{n=0}^N}' g(i \xi_n)} &<& u \ll 1~.
\end{eqnarray}
To avoid the evaluation of the integral in Eq.(\ref{eq:n_numerik}) we have used an upper bound 
\begin{equation}
 \frac{N g(i \xi_N) /\tau}{{\sum_{n=0}^N}' g(i \xi_n)} < u~, \label{eq:n_numerik2}
\end{equation}
which exploits the exponential decay of $g$ introduced by the Green's tensor. Typically, $\tau \sim L/\Lambda_T$.
In our numerical calculations, we targeted errors $u \in [10^{-3}, 10^{-6}]$ and obtained sums over $N \in [10^2,10^5]$ terms.
In some cases like BCS theory, the calculation of the optical response is not trivial and the remainder integral was neglected completely. This requires a small enough value of $u$ and yields a systematic numeric error $\mathcal{O}(u)$.

\section{Magnetic transition matrix elements}\label{Magnetic transition matrix elements}
This section gives the magnetic transition matrix elements used in the calculations of polarizabilities and free energies according to Eqs.(\ref{WS}, \ref{eq:WS_polarizability}) or \eqref{eq:WS_polarizability_thermal}.

The simplest approach used is the spin $\frac{1}{2}$ system (two-level atom) with states $|m_S\rangle = |\pm\frac{1}{2}\rangle$.
The dipole operator is $\boldsymbol{\mu} = \mu_B \hbar g_S\boldsymbol{S}$ and the quantization axis is $\hat z$,
so that
\begin{eqnarray}
S_x|m_S\rangle &=& \frac{1}{2} |-m_S\rangle \label{eq:spin_OP_x}~,\\
S_y|m_S\rangle &=& i m_S |-m_S\rangle\label{eq:spin_OP_y}~,\\
S_z|m_S\rangle &=& m_S |m_S\rangle\label{eq:spin_OP_z}~.
\end{eqnarray}
According to Eq. (\ref{eq:two-level_polarizability}) we need only matrix elements connecting to the ground-state $|g\rangle = |-\frac{1}{2}\rangle$, the only nonvanishing of which yield
\begin{eqnarray}\textstyle
|\langle g  | S_x |e\rangle|^2 = |\langle g | S_y |e\rangle|^2  =  |\langle g| S_z |g \rangle|^2 =  \frac{1}{4}.
\end{eqnarray}
Since all polarizabilities \eqref{WS}, \eqref{eq:WS_polarizability} or \eqref{eq:WS_polarizability_thermal} are proportional to the transition frequency, the $S_z$ transitions connecting identical states do not contribute and we obtain the result of Eq. \eqref{eq:FE-aniso}.
%
%

Let us now calculate the matrix elements for the $\mathrm{^{87}Rb}$ atom
prepared in a hyperfine s-state, 
\begin{equation}
\mu_i^{ab} = \langle a | \mu_i | b\rangle=\langle F, m_F | \mu_i | F', m_F'\rangle~.
\end{equation}
The dipole operator $\boldsymbol{\mu} = \mu_B \hbar (g_S \boldsymbol{S} + g_L \boldsymbol{L} + g_I \boldsymbol{I}) \approx \mu_B \hbar g_S \boldsymbol{S}$, assuming vanishing orbital momentum and because of the smallness of the nuclear Land\'{e} g-factor $g_I \propto m_e/m_p$.
We express the states $|a\rangle, |b\rangle$ in the uncoupled basis of the spin and nuclear spin momenta with the help of the Clebsch-Gordan coefficients
\begin{equation}
|F,m_F\rangle = \sum_{m_I, m_S} C^{m_I, m_S}_{F,m_F} |m_I, m_S\rangle~,
\end{equation}
where the action of the components of the spin operator on a state is known [Eqs.(\ref{eq:spin_OP_x}-\ref{eq:spin_OP_z})].
Hence, we find
\begin{eqnarray}
\mu_x^{ab} &=&  g_S \mu_B \sum_{m_I, m_S} \frac{1}{2} C^{m_I, -m_S}_{F',m_F'} C^{m_I, m_S}_{F,m_F}~,\\
\mu_y^{ab} &=& - i g_S \mu_B \sum_{m_I, m_S} m_S C^{m_I, -m_S}_{F',m_F'} C^{m_I, m_S}_{F,m_F}~,\\
\mu_z^{ab} &=& - g_S \mu_B \sum_{m_I, m_S} m_S C^{m_I, m_S}_{F',m_F'} C^{m_I, m_S}_{F,m_F}~.
\end{eqnarray}
%


\end{document}